\documentclass[12pt]{article}
\usepackage{float}
\usepackage{amssymb, amsmath, amsthm}
\usepackage{mathrsfs}
\usepackage{tabulary}
\usepackage{rotating}
\usepackage{enumerate}
\usepackage{setspace}
\usepackage{multirow}
\usepackage{paralist}
\usepackage{listings}
\usepackage[T1]{fontenc}
\usepackage[utf8]{inputenc}
\usepackage[english]{babel}
\usepackage{authblk}
\usepackage{etoolbox}
\usepackage{mathtools}
\usepackage{longtable}
\usepackage[numbers,round,sort&compress]{natbib}
\usepackage[usenames,dvipsnames]{color}
\usepackage{bibentry}

\newcommand\BibTeX{{\rmfamily B\kern-.05em \textsc{i\kern-.025em b}\kern-.08em
T\kern-.1667em\lower.7ex\hbox{E}\kern-.125emX}}

\usepackage{caption}

\IfFileExists{upquote.sty}{\usepackage{upquote}}{}

\usepackage{hyperref}
\usepackage{subcaption}
\newtheorem{thm}{Theorem}[section]

\theoremstyle{definition}

\usepackage[hmargin=2.54cm,vmargin=2.54cm]{geometry}

\title{Transporting stochastic direct and indirect effects to new populations}
\author[1]{Kara E. Rudolph\thanks{Corresponding author: \\ 2315 Stockton Blvd. Sacramento, CA 95817\\
kerudolph@ucdavis.edu \\ tel. +15107619404 }}
\author[2]{Jonathan Levy}
\author[2]{Mark J. van der Laan}
\affil[1]{\footnotesize Department of Emergency Medicine, School of Medicine, University of California, Davis, Sacramento, California}
\affil[2]{\footnotesize Division of Biostatistics, School of Public Health, University of California, Berkeley, California}
\date{}

\renewcommand{\refname}{References}

\begin{document}
\maketitle

\begin{abstract}
Transported mediation effects may contribute to understanding how and why interventions may work differently when applied to new populations. However, we are not aware of any estimators for such effects. Thus, we propose several different estimators of transported stochastic direct and indirect effects: an inverse-probability of treatment stabilized weighted estimator, a doubly robust estimator that solves the estimating equation, and a doubly robust substitution estimator in the targeted minimum loss-based framework. We demonstrate their finite sample properties in a simulation study.
\end{abstract}

\noindent Keywords: Causal inference; Mediation; Transportability; Generalizability; External validity; Instrumental variables; Targeted maximum likelihood estimation
\section{Introduction}

Often, an intervention, program, or policy that works in one place or population fails to replicate in another place or population \citep{rudolph2018composition} or can even have unintended harmful effects \citep{kling2007experimental}. This is problematic from a public policy or public health perspective in that the goals of such interventions are to help---not harm, and problematic from a financial perspective in that limited resources may be not be spent optimally. 

When such initiatives fail to replicate or have unintended effects in new populations, transportability theory and methods offer a chance to understand why. Transportability is the ability (based on identifying assumptions) to transport a causal effect from a source population to a new, target population, accounting for differences between the two populations (e.g., differences in compositional factors, treatment adherence, etc.) \citep{pearl2014external}. Previous work developed estimators to transport total effects from a source to target population \citep{rudolph2016robust} or, similarly, to generalize effects from a sample to the population \citep{miettinen1972standardization,stuart2011use, cole2010generalizing,frangakis2009calibration}. 

In some cases, examining transportability of the total effect may shed light on reasons for lack of replication. However, in other cases, transporting the total effect may not identify the relevant differences and it may be beneficial to go further and examine transportability of the underlying mediation mechanisms. 
 Although there has been work on the identification on transported indirect effects \citep{bareinboim2013general,pearl2014external}, we are not aware of any previous work developing estimators for transporting mediation effects (direct and indirect effects) from a source to target population.
 Thus, we address this research gap by proposing several different estimators of stochastic direct and indirect effects: an inverse-probability of treatment stabilized weighted estimator, a doubly robust estimator that solves the estimating equation, and a doubly robust substitution estimator in the targeted minimum loss-based framework.  

The paper is organized as follows. In Section 2, we introduce notation and the structural equations model generating our data. In Section 3, we define the parameters of interest, the transported stochastic direct and indirect effects, and give identification results. Sections 4, 5, and 6, detail an inverse-probability of treatment weighted estimator, an estimating equation estimator, and a targeted minimum loss-based estimator for estimating these transported stochastic direct and indirect effects. In Section 7, we present the results of simulation studies that demonstrates the relative performance of the aforementioned estimators in finite samples. Section 8 concludes. 

\section{Notation and structural equations model}
The full data is generated by a structural equations model (SEM) \citep{wright1921correlation,pearl2009causal}, which consists of the data generating process to which we would like to have access. The SEM first generates a random draw of a vector $U$ of unknown measurements \citep{pearl2009causality}, where $U=(U_{S},U_{W},U_{A},U_{Z},U_{M},U_{Y})\sim P_U$. Then our variables are generated in the following time ordering: 
\begin{eqnarray*}
S & = & f_{S}(U_{S})\\
W & = & f_{W}(U_{W},S)\\
A & = & f_{A}(U_{A},W,S)\\
Z & = & f_{Z}(U_{Z},A,W,S)\\
M & = & f_{M}(U_{M},Z,W,S)\\
Y & = & f_{Y}(U_{Y},Z,W,M),
\end{eqnarray*} 
where $S$ is a binary indicator of site, $W$ is a vector of covariates, $A$ is a binary treatment, $Z$ is a binary intermediate variable, $M$ is a binary mediator, and $Y$ is a binary or continuous outcome. The SEM generates the full data as $(U,O)\sim P_{UO} \in \cal{M}^F$, our full-data statistical model.  If we had access to the SEM, we could generate potential (i.e., counterfactual) outcomes \citep{neyman1923application,rubin1974estimating}, which define our causal parameters of interest. We observe data $O=(S, W, A, Z, M, S\times Y)$ for $n$ participants, with the true distribution $O_1, ..., O_n \overset{iid}{\sim} P_O \in \cal{M}$, our observed data statistical model.  Note, we will only observe the outcome, $Y$, for site $S=1$.  

We consider a structural causal model for these data aligned with our motivating example in which $A$ is an instrumental variable (IV) \citep{angrist1996identification}, which puts several restrictions on the statistical model, $\cal{M}$, where $\cal{M}$ is the collection of probability distributions under which the causal quantities of interest are identified---we define and identify such quantities in the following section. The restrictions are: 1) $A$ is randomly assigned (possibly conditional on $(W,S)$), and 2) there is no direct effect of $A$ on $M$ or of $A$ on $Y$---downstream effects of $A$ only operate through $Z$ \citep{rudolph2017robust}. However, the methods we propose can be used in the more general case where $A$ directly affects $M$ and/or $Y$; we discuss such an extension in the appendix. 

%
%

In our notation, we use lowercase letters to denote fixed, assignment values of variables and uppercase letters to denote observed values.  We use subscripts for descriptive purposes---subscripts are not to be considered a variable. For instance, we use a capital letter in $p_Y$, the conditional density of $Y$, because it is a density of the random variable $Y$. 
 
\section{Parameters of interest}

We consider two causal quantities of interest that we call transported stochastic direct and indirect effects. These causal quantities represent stochastic direct and indirect effects \citep{rudolph2017robust,vanderweele2016mediation} transported from a source population to a new, target population. Stochastic direct and indirect effects, also called randomized interventional direct and indirect effects \citep{vanderweele2016mediation}, represent the 1) direct effect of $A$ on $Y$ not through $M$ and the 2) indirect effect of $A$ on $Y$ through $M$. As has been described previously \citep{rudolph2018causal}, one can consider versions of these effects that condition on $Z$ and thus estimate the indirect pathway of $A$ to $M$ to $Y$, not through $Z$ \citep{zheng2017longitudinal}, or versions that marginalize over $Z$ and thus estimate the combined indirect pathways of 1) $A$ to $Z$ to $M$ to $Y$ plus 2) $A$ to $M$ to $Y$ \citep{rudolph2017robust,vanderweele2016mediation}. Adhering to the IV constraints on our statistical model, no effect operates through pathway $A$ to $M$ to $Y$, so we focus herein on the versions of these effects that marginalize over $Z$. Previously, such a stochastic intervention on $M$ has been defined \citep{rudolph2017robust, vanderweele2016mediation}
\[
g^*_{M\mid a^{*},W}(M\mid W)=\sum_{z}Pr(M\mid Z=z,W)Pr(Z=z\mid A=a^{*},W).
\]
The subscript for $\hat{g}_{M\mid a^*, W}$ specifies that it is a conditional density of random variable $M$ given random variable $W$, and value $a^*$ for which a lower case letter indicates they are fixed and the same for all participants. Previously, the parameter of interest was defined as $\Psi^F(P_{UO})=\mathbb{E}\left[Y_{a,g^*_{M\mid a^{*},W}}\right]$
where the expectation is taken over the full data model and $Y_{a,g
^*_{M\mid a^{*},W}}$
is a potential outcome intervening on $A$ to set it to $a$, and then downstream, intervening on $M$ to set it to a random (i.e., stochastic) draw from the distribution of $M$ defined by $g^*_{M\mid a^{*},W}(M | W)$ \citep{rudolph2017robust}.   We wish to transport this parameter to a new site
where the outcome was not observed $(S=0)$, and thus make the following modification:
\[
\Psi^F(P_{UO})=\mathbb{E}\left[Y_{a,g^*_{M\mid a^{*},W, s}}\mid S=0\right]
\]
 where 
\begin{equation}
g^*_{M\mid a^{*},W, s}(M \mid W)=\sum_{z}Pr(M\mid Z=z,W,S=s)Pr(Z=z\mid A=a^{*},W,S=s),
\end{equation}
and where we impose a certain $a^*$ and a certain $s$ in both the $Z$ and $M$ models. We note that $g^*_{M\mid a^{*},W, s}(M|W)$ represents any stochastic intervention, which can include stochastic draws from the true models, but can also include a data-dependent version, estimated from observed data distributions, which we denote $\hat{g}_{M\mid a^{*},W, s}(M|W)$. 

The transported stochastic direct effect entails setting $a^*$ to 0 and taking the difference in mean outcome between setting $a$ to 1 and setting $a$ to 0, denoted 
\[ 
\mathbb{E}\left[Y_{1,g^*_{M\mid 0,W, s}} - Y_{0,g^*_{M\mid 0,W, s}}\mid S=0\right] 
\] 
and the transported stochastic indirect effect entails setting $a=1$ and then taking the difference in mean outcome between setting $a^*=1$ and $a^*=0$, denoted 
\[ 
\mathbb{E}\left[Y_{1,g^*_{M\mid 1,W, s}} - Y_{1,g^*_{M\mid 0,W, s}}\mid S=0\right].
\] 

\section{Identifiability}
\label{idSDE}
To identify the stochastic direct effect and stochastic indirect effect we will need to impose additional assumptions on $\mathcal{M}^F$ and $\mathcal{M}$, listed below.  

\begin{enumerate}
\item Positivity: For all $S$ and $W$ we need a positive probability of
assigning any level of $A$. For all combinations of $S, W,$ and $A=a$, we have a positive probability of any level of $Z.$ For
$S=1$ and all combinations of $Z$ and $W$ we need a positive probability
of any level of the mediator, $M$.  

\item Common outcome model across sites: $\mathbb{E}\left[Y\mid M,Z,W,S=1\right]=\mathbb{E}\left[Y\mid M,Z,W,S=0\right]$. The null hypothesis of a common outcome model may be tested nonparametrically \citep{luedtke2015omnibus}.

\item Sequential Randomization: $Y_{am}\perp A\mid W,S$ and $Y_{am}\perp M\mid W,Z,S$.  
This is akin to a two-time point longitudinal intervention where at the first time point, we statically intervene to set the treatment, $A=a$, and at the second time point, we stochastically intervene on the mediator, $M$.  
\end{enumerate}


\begin{thm}
Given the above assumptions, we can establish the following identification:
\begin{align*}
& \Psi(P) = \Psi^F(P_{UX}) = \mathbb{E}\biggr[\mathbb{E}\biggr[\mathbb{E}_{\hat{g}_{M \mid a^*,W,s}}\biggr[\mathbb{E}\biggr[Y | W,Z,M, S=1 \biggr] | W,Z\biggr] | W,a,S=0\biggr] | S=0\biggr] \\
&= \mathbb{E} \biggr[ \mathbb{E}\biggr[\sum_m \biggr[  \mathbb{E}Y\hat{g}_{M\mid a^*,W,s}(m \mid W) \mid M=m, W,Z,A=a, S =1 \biggr] \mid A=a, W,S \biggr] \mid S = 0 \biggr]
\end{align*}
\end{thm}
\noindent The proof is in the appendix.

\section{Stabilized Inverse probability of treatment weighted estimator}
First, we describe a stabilized inverse probability of treatment weighted (IPTW) estimator of $\Psi(P)$. The R code to implement this estimator is provided in the appendix. We describe model fitting using regression language for simplicity but note that machine learning can be used instead. We will use the knowledge of our smaller model, $\cal{M}$, with restrictions detailed in Section 2, so do not include $A$ in the regression model for $M$.

We use the weights \begin{multline}
\label{weights}
H(M,Z,A,W,S)= \\
\frac{\hat{g}_{M\mid a^{*},W,s}(M \mid W)p_{Z}\left(Z\mid A=a,W,S=0\right)p_{W}\left(S=0\mid W\right)I(S=1,A=a)}{p_{M}\left(M\mid Z,W,S=1\right)p_{Z}\left(Z\mid A=a,W,S=1\right)p_{A}\left(a\mid W,S=1\right)p_{W}\left(S=1\mid W\right)P_{S}(S=0)}.
\end{multline}
$\hat{g}_{M|a^*,W,s}(M|W)$ is a data-dependent stochastic intervention on $M$, which can be estimated $\hat{g}_{M|a^*,W,s}(M|W) = \sum^1_{z=0}P(M=1 | Z=z,W, S=s) P(Z=z | A=a^*, W, S=s).$ $P(M=1 | Z=z,W, S=s)$ can be estimated using a logistic regression estimating the probability of $M=1$ given $Z$, $W$, and $S$ and thereby getting predicted probabilities for $M=1$ setting $S=s$ and separately setting $Z=1$ and $Z=0$. $P(Z=z | A=a^*, W, S=s)$ can be estimated using a logistic regression estimating the probability of $Z=1$ given $A$, $W$, and $S$ and thereby getting predicted probabilities for $Z=1$ and for $Z=0$, setting $A=a^*$ and $S=s$ and using observed values for $W$. We note that marginalizing out $Z$ introduces dependence on $A$.  

The stabilized IPTW estimate of $\Psi(P)$ is the empirical mean of $Y$ weighted by $\hat{H}_n(M,Z,A,W,S)$, stabilized by the empirical mean of $\hat{H}(M,Z,A,W,S)$: 
\begin{equation}
 \hat{\Psi}_n^{IPTW}= \frac{\frac{1}{n}\sum_i\hat{H}_n(M_i,Z_i,A_i,W_i,S_i)Y_i}{\frac{1}{n}\sum_i\hat{H}_n(M_i,Z_i,A_i,W_i,S_i)},
\end{equation}
where $\hat{H}_n$ is an approximation of $H(M,Z,A,W,S)$.

The standard error of the stabilized IPTW estimator is calculated as the sample standard deviation of the approximation of the IPTW influence curve of $\Psi(P)$, given by $$IC(P) = \frac{\hat{H}_n(M_i,Z_i,A_i,W_i,S_i)}{\frac{1}{n}\sum_i\hat{H}_n(M_i,Z_i,A_i,W_i,S_i)}\bigg(Y_i - \frac{1}{n}\sum_i\hat{H}_n(M_i,Z_i,A_i,W_i,S_i)Y_i\bigg).$$ 

The stochastic direct effect (SDE) entails setting $a^*$ to 0 and taking the difference in estimates between setting $a$ to 1 and setting $a$ to 0. The corresponding influence curve approximation is a difference of the influence curve approximation for the parameter defined by setting $a^* = 0, a=1$ and the influence curve approximation for the parameter defined by setting $a^* = 0, a=0$. The stochastic indirect effect (SIE) entails setting $a=1$ and then taking the difference in estimates between setting $a^*=1$ and $a^*=0$. The corresponding influence curve approximation is again a difference of the two influence curve approximations (one for parameter defined by setting $a^* = 1, a=1$ and the other for $a^* = 0, a=1$). For each estimator, we used the sample standard deviation of their respective influence curve approximations, divided by $\sqrt{n}$ for the standard error estimate.  




\section{Estimating equation estimator}
Next, we describe two estimating equation (EE) estimators of $\Psi(P)$: 1) one that incorporates the exclusion restrictions on our statistical model, $\cal{M}$, that there is no direct effect of $A$ on $M$ or of $A$ on $Y$, and 2) another that does not impose those restrictions. The R code to implement these estimators is provided in the appendix. We describe model fitting using regression language for simplicity but note that machine learning can be used instead. 

\subsection{Estimator incorporating exclusion restrictions}
This EE estimator solves the efficient influence curve (EIC) equation of the parameter $\Psi(P)$ for the restricted model where $M$ and $Y$ do not depend directly on $A$. This EIC is given by 
\begin{align}
\begin{split}
\label{eicrestriction}
D(P) &= D_Y(P)+D_Z(P)+D_W(P), \text{ where }\\
D_{Y}(P)&=\left(Y-\bar{Q}_{Y}(M,Z,W)\right)\\
&\times \frac{\hat{g}_{M\mid a^{*},W,s}(M|W)p_{Z}\left(Z\mid A=a,W,S=0\right)p_{W}\left(S=0\mid W\right)I(S=1)}{g_{M}\left(M\mid Z,W,S=1\right)p_{Z}\left(Z\mid W,S=1\right)p_{W}\left(S=1\mid W\right)P_{S}(S=0)}, \\
D_{Z}(P) &= \left(\bar{Q}_{M}(Z,W,S)-\bar{Q}_{Z}(a, W, S)\right)\frac{I(S=0,A=a)}{p_{A}\left(a\mid W,S\right)p_{S}(S=0)}, \text{ and }\\
D_{W}(P) &= \left(\bar{Q}_{Z}(a, W, S)-\Psi(P)\right)\frac{I(S=0)}{p_{S}(S=0)} \text{ (notation is explained further below}).
\end{split}
\end{align}

We reiterate that this EIC is only efficient for model with the exclusion restrictions. We approximate $D_Y(P)$ as follows. First, we estimate $\mathbb{E}\left[Y\mid M,Z,W\right]$ as $\bar{Q}_{Y,n}(M,Z,W),$ which can be calculated using predicted values from a regression of $Y$ on $M, Z, W$ among those with $S=1$. $p_Z(Z | W,S=1)$ can be calculated $p_Z(Z | W,S=1) = \sum_{a=0}^1 P(Z=1 | A=a, W, S=1)P(A=a | W, S=1)$. Each of the remaining probabilities in $D_Y(P)$ can be calculated using predicted probabilities from a logistic regression of the relevant conditional mean outcome models for M, Z and S, as described in the stabilized IPTW estimator section. Estimation of  $\hat{g}_{M|a^*,W,s}(M|W)$ is also described in the stabilized IPTW estimator section.

We next approximate $D_Z(P)$. To do so, we apply the stochastic intervention on $\bar{Q}_{Y,n}(M,Z,W)$
via the computation $\bar{Q}_{M,n}(Z,W,S)=\mathbb{E}_{\hat{g}_{M\mid a^{*},W,s}}[\bar{Q}_{Y,n}(M,Z,W) \mid Z,W,S]$. 
 Specifically, we can generate predicted values of $\bar{Q}_{Y,n}(1,Z,W)$ and $\bar{Q}_{Y,n}(0,Z,W)$ and then marginalize over $\hat{g}_{m\mid a^{*},W,s}(M|W)$: $\sum^1_{m=0} \bar{Q}_{Y,n}(m,Z,W)\hat{g}_{m\mid a^{*},W,s}(m|W)$. Next, we estimate $\bar{Q}_{Z,n}(a,W,S)$ by regressing $\bar{Q}_{M,n}(Z,W,S)$ on $A, W, $ and $S$, and getting predicted values setting $A=a$. 

The estimate of $\Psi(P)$ is obtained by solving $\frac{1}{n}\sum_i \hat{D}_n(O_i) = 0$, where $\hat{D}_{\Psi,n}$ is the EIC approximation as computed above.  
The variance of the EE estimate is the sample variance of the EIC approximation. The SDE, SIE, and their corresponding ICs and standard errors can be calculated as described in the stabilized IPTW estimator section. 
  The conditions for consistency and asymptotic efficiency of this estimator are discussed in the appendix.  
\subsection{Estimator not imposing exclusion restrictions}
This EE estimator solves the efficient influence curve (EIC) equation of the parameter $\Psi(P)$ for the unrestricted model where $M$ and $Y$ may depend directly on $A$. It will be inefficient under the restricted model where $M$ and $Y$ do not depend directly on $A$. This EIC is given by 
\begin{align}
\begin{split}
\label{eicnorestriction}
D(P) &= D_Y(P)+D_Z(P)+D_W(P), \text{ where }\\
D_{Y}(P)&=\left(Y-\bar{Q}_{Y}(M,Z,A,W)\right)\\
&\times \frac{\hat{g}_{M\mid a^{*},W,s}(M|W)p_{Z}\left(Z\mid A=a,W,S=0\right)p_{W}\left(S=0\mid W\right)I(S=1,A=a)}{g_{M}\left(M\mid Z,A,W,S=1\right)p_{Z}\left(Z\mid A=a,W,S=1\right)p_{A}\left(a\mid W,S=1\right)p_{W}\left(S=1\mid W\right)P_{S}(S=0)}, \\
D_{Z}(P) &= \left(\bar{Q}_{M}(Z,A,W,S)-\bar{Q}_{Z}(a, W, S)\right)\frac{I(S=0,A=a)}{p_{A}\left(a\mid W,S\right)p_{S}(S=0)}, \text{ and }\\
D_{W}(P) &= \left(\bar{Q}_{Z}(a, W, S)-\Psi(P)\right)\frac{I(S=0)}{p_{S}(S=0)} \text{ (notation is explained further below}).
\end{split}
\end{align}

We approximate $D_Y(P)$ as follows. First, we estimate $\mathbb{E}\left[Y\mid M,Z,A,W\right]$ as $\bar{Q}_{Y,n}(M,Z,A,W),$ which can be calculated using predicted values from a regression of $Y$ on $M, Z, A, W$ among those with $S=1$. Each of the probabilities in $D_Y(P)$ can be calculated using predicted probabilities from a logistic regression of the relevant conditional mean outcome models for M, Z, A, and S. $\hat{g}_{M|a^*,W,s}(M|W)$ can be estimated $\hat{g}_{M|a^*,W,s}(M|W) = \sum^1_{z=0}P(M=1 | Z=z, A=a^*, W, S=s) P(Z=z | A=a^*, W, S=s).$   

We next approximate $D_Z(P)$. To do so, we apply the stochastic intervention on $\bar{Q}_{Y,n}(M,Z,A,W)$
via the computation $\bar{Q}_{M,n}(Z,A,W,S)=\mathbb{E}_{\hat{g}_{M\mid a^{*},W,s}}[\bar{Q}_{Y,n}(M,Z,A,W) \mid Z,W,S]$. 
 Specifically, we can generate predicted values of $\bar{Q}_{Y,n}(1,Z,A,W)$ and $\bar{Q}_{Y,n}(0,Z,A,W)$ and then marginalize over $\hat{g}_{m\mid a^{*},W,s}(M|W)$: $\sum^1_{m=0} \bar{Q}_{Y,n}(m,Z,A,W)\hat{g}_{m\mid a^{*},W,s}(m|W)$. Next, we estimate $\bar{Q}_{Z,n}(a,W,S)$ by regressing $\bar{Q}_{M,n}(Z,A,W,S)$ on $A, W, $ and $S$, and getting predicted values setting $A=a$. 

The estimate of $\Psi(P)$ is obtained by solving $\frac{1}{n}\sum_i \hat{D}_n(O_i) = 0$, where $\hat{D}_{\Psi,n}$ is the EIC approximation as computed above.  
 The variance of the EE estimate is the sample variance of the EIC approximation. 
  The SDE, SIE, and their corresponding ICs and standard errors can be calculated as described in the stabilized IPTW estimator section. The conditions for consistency and asymptotic efficiency of this estimator are discussed in the appendix.
  
\section{Targeted minimum loss-based estimator}
We now describe how to estimate $\Psi(P)$ using targeted minimum loss-based estimation (TMLE). This estimation approach, which is just one of several TMLE approaches that could be used, uses sequential regression, updating the conditional outcome model at each stage to both solve the EIC equation while also lowering the empirical negative log-likelihood loss.  The process is identical to a two time-point longitudinal intervention \citep{van2012targeted}. Similar to the EE estimator section, we describe two TMLE estimators of $\Psi(P)$: 1) one that incorporates the exclusion restrictions on our statistical model, $\cal{M}$, that there is no direct effect of $A$ on $M$ or of $A$ on $Y$, and 2) another that does not impose those restrictions. We include R code for these estimators in the appendix. As in the previous sections, we describe model fitting using regression language for simplicity but note that machine learning can be used instead. 

\subsection{Estimator incorporating exclusion restrictions}
This TMLE estimator solves the EIC equation of the parameter $\Psi(P)$ for the restricted model where $M$ and $Y$ do not depend directly on $A$.

Let $\bar{Q}_{Y,n}^{0}(M,Z,W)$ be an initial estimate of $\mathbb{E}\left[Y\mid M,Z,W\right]$. 
$\bar{Q}_{Y,n}^{0}(M,Z,W)$ can be estimated by predicted values from a regression of $Y$ on $M, Z, W$ among those with $S=1$. 

Next, we update that initial estimate using the weights
\begin{multline}
\label{weightsnoa}
H(M,Z,W,S)=
\frac{\hat{g}_{M\mid a^{*},W,s}(M|W)p_{Z}\left(Z\mid A=a,W,S=0\right)p_{W}\left(S=0\mid W\right)I(S=1)}{g_{M}\left(M\mid Z,W,S=1\right)p_{Z}\left(Z\mid W,S=1\right)p_{W}\left(S=1\mid W\right)P_{S}(S=0)},
\end{multline}
which are approximated with $\hat{H}_n(M,Z,W,S).$ 
$\bar{Q}_{Y,n}^{0}(M,Z,W)$ is updated by performing a weighted parametric logistic regression of $Y$ with $logit(\bar{Q}_{Y,n}^{0}(M,Z,W))$ as an offset, intercept $\epsilon_Y$, and weights $\hat{H}_n(M,Z,W,S)$. $\epsilon_{Y,n}$ is the MLE fit of intercept $\epsilon_Y$. The update is given by $\bar{Q}_{Y,n}^{*}(M,Z,W)= \bar{Q}_{Y,n}^{0}(\epsilon_{Y,n})(M,Z,W)$. 

We then perform the stochastic intervention on $\bar{Q}_{Y,n}^{*}(M,Z,W)$
via the computation $\bar{Q}_{M,n}^{*}(Z,W,S)=\mathbb{E}_{\hat{g}_{M\mid a^{*},W,s}}[\bar{Q}_{Y,n}^{*}(M,Z,W) \mid Z,W,S]$. This can be done by generating predicted values of $\bar{Q}_{Y,n}^{*}(1,Z,W)$ and $\bar{Q}_{Y,n}^{*}(0,Z,W)$ and then marginalizing over $\hat{g}_{m\mid a^{*},W,s}(M|W)$: $\sum^1_{m=0} \bar{Q}_{Y,n}^{*}(m,Z,W)\hat{g}_{m\mid a^{*},W,s}(m|W)$. 

Next, we estimate $\bar{Q}_{Z,n}^{0}(a,W,S)$ by regressing $\bar{Q}_{M,n}^{*}(Z,W,S)$ on $A, W, $ and $S$, and getting predicted values setting $A=a$. We then update this initial estimate using a second set of weights, 
\begin{equation}
H_{a}(a,W,S)=\frac{I(S=0,A=a)}{p_{A}\left(A\mid W,S\right)p_{S}(S=0)},
\end{equation} 
in a weighted logistic regression of $logit(\bar{Q}_{M, n}^{*}(Z,W,S))$ with $logit(\bar{Q}_{Z,n}^{0}(a,W,S))$ as an offset, intercept $\epsilon_Z$. $\epsilon_{Z,n}$ is the MLE fit of intercept $\epsilon_Z$. The updated estimate will be notated $\bar{Q}_{Z,n}^{*}(a,W,S) = \bar{Q}_{Z,n}^{0}(\epsilon_{Z,n})(a,W,S)$.

The empirical mean of $\bar{Q}_{Z,n}^{*}(a,W,S)$ among those for whom $S=0$ is the TMLE estimate of $\Psi(P)$. The influence curve approximation obtained by incorporating our updated model fits solves the empirical mean of the EIC equation (see Equation \ref{eicrestriction} in Section 6), replacing $P$ with $P^*_n$ defined by $\bar{Q}_{Y,n}^{*}(M,Z,W))$, $\bar{Q}_{M,n}^{*}(Z,W,S)$, $\bar{Q}_{Z,n}^{*}(a,W,S)$, our known $p_A(A|W,S)$ and $p_S(S=0)$, our regression fit for $p_M(M|Z,W,S), p_Z(Z|A,W,S)$ and $p_{S|W}(S|W)$,  and the empirical distribution as an estimate of $p_W(W)$. We denote this EIC $D(P^*_n)$. The TMLE updating step also decreases the empirical loss of the model fits. The variance of the TMLE estimate is the sample variance of the EIC approximation. The SDE, SIE, and their corresponding ICs and standard errors can be calculated as described in the stabilized IPTW estimator section. The conditions for consistency and asymptotic efficiency of this estimator are discussed in the appendix.

\subsection{Estimator not imposing exclusion restrictions}
This TMLE estimator solves the EIC equation of the parameter $\Psi(P)$ for the unrestricted model where $M$ and $Y$ may depend directly on $A$. It will be inefficient under the restricted model where $M$ and $Y$ do not depend directly on $A$.

Let $\bar{Q}_{Y,n}^{0}(M,Z,A,W)$ be an initial estimate of $\mathbb{E}\left[Y\mid M,Z,A,W\right]$. 
$\bar{Q}_{Y,n}^{0}(M,Z,A,W)$ can be estimated by predicted values from a regression of $Y$ on $M, Z,A, W$ among those with $S=1$. 

Next, we update that initial estimate using the weights, $H(M,Z,A,W,S)$, given in Equation \ref{weights}, which are approximated with $\hat{H}_n(M,Z,A,W,S).$ 
$\bar{Q}_{Y,n}^{0}(M,Z,A,W)$ is updated by performing a weighted parametric logistic regression of $Y$ with $logit(\bar{Q}_{Y,n}^{0}(M,Z,A,W))$ as an offset, intercept $\epsilon_Y$, and weights $\hat{H}_n(M,Z,A,W,S)$. $\epsilon_{Y,n}$ is the MLE fit of intercept $\epsilon_Y$. The update is given by $\bar{Q}_{Y,n}^{*}(M,Z,A,W)= \bar{Q}_{Y,n}^{0}(\epsilon_{Y,n})(M,Z,A,W)$. 

We then perform the stochastic intervention on $\bar{Q}_{Y,n}^{*}(M,Z,A,W)$
via the computation $\bar{Q}_{M,n}^{*}(Z,A,W,S)=\mathbb{E}_{\hat{g}_{M\mid a^{*},W,s}}[\bar{Q}_{Y,n}^{*}(M,Z,A,W) \mid Z,A,W,S]$. This can be done by generating predicted values of $\bar{Q}_{Y,n}^{*}(1,Z,A,W)$ and $\bar{Q}_{Y,n}^{*}(0,Z,A,W)$ and then marginalizing over $\hat{g}_{m\mid a^{*},W,s}(M|W)$: $\sum^1_{m=0} \bar{Q}_{Y,n}^{*}(m,Z,A,W)\hat{g}_{m\mid a^{*},W,s}(m|W)$. 

Next, we estimate $\bar{Q}_{Z,n}^{0}(a,W,S)$ by regressing $\bar{Q}_{M,n}^{*}(Z,A,W,S)$ on $A, W, $ and $S$, and getting predicted values setting $A=a$. The remainder of the steps for this TMLE are identical to those for the restricted TMLE in the above subsection. 

\section{Simulation}
\subsection{Overview}
We compare finite sample performance of our three types estimators in estimating the transported SDE and transported SIE using simulation. For both TMLE and EE, we include the 1) versions that are efficient under the restricted model in which $M$ and $Y$ do not depend directly on $A$ (henceforth TMLE efficient and EE efficient) and 2) versions that are more general in that they allow for the exclusion restriction not to hold but are inefficient under resetricted model $\cal{M}$ (henceforth TMLE general and EE general). We show estimator performance in terms of absolute bias, efficiency, 95\% confidence interval (CI) coverage, root mean squared error (RMSE), and percent of estimates lying outside the bounds of the parameter space across 1,000 simulations. For calculating the efficiency and the 95\% CI coverage, we use both the IC and 500 bootstrap replicates.

We consider two data-generating mechanisms (DGMs) within the structural causal model described in Section 2. The DGMs are detailed in Table \ref{dgmtab} using the same notation as in Section 2. 
\begin{table}[!h]
\centering
\footnotesize
\caption{Simulation data-generating mechanisms.}
\label{dgmtab}
\begin{tabular}{| p{2.5cm} | p{13.0cm} | }
  \hline
\multicolumn{2}{|c|}{Data Generating Mechanism 1}\\
\hline 
$W_1 \sim bernoulli $ & $P(W_1 = 1) = 0.5$ \\
$W_2 \sim bernoulli $ & $P(W_2 = 1) = expit(0.4 + 0.2W_1)$\\
$S \sim bernoulli $& $P(S = 1) = expit(3W_2 - 1)$\\
$A \sim bernoulli $& $P(A = 1) = 0.5 $ \\
$Z \sim bernoulli $& $P(Z = 1) = expit(-0.1 A + -0.2 S + 0.2 W_2 + 5 AW_2 + 
        0.14 A S + 0.2 W_2 S - 0.2 A W_2 S - 1)$\\
$M \sim bernoulli $& $P(M = 1) = expit(1Z + 3ZW_2 + 0.2ZS - 0.2
        W_2S + 2W_2Z + 0.2S - 0.2ZW_2S - W_2 - 
        2)$\\
$Y \sim bernoulli $& $P(Y = 1) = expit(-6Z + 0.2ZW_2 + 2ZM + 2W_2
        M - 2W_2 + 4M + 1ZW_2M - 0.2)$\\
        \hline
\multicolumn{2}{|c|}{Data Generating Mechanism 2}\\       
\hline
  $W_1 \sim bernoulli $ & $P(W_1 = 1) = 0.5$ \\
$W_2 \sim bernoulli $ & $P(W_2 = 1) = expit(0.4 + 0.2W_1)$\\
$S \sim bernoulli $& $P(S = 1) = expit(3W_2 - 1)$\\
$A \sim bernoulli $& $P(A = 1) = 0.5 $ \\
$Z \sim bernoulli $& $P(Z = 1) = expit(-3A + -0.2S + 2W_2 + 0.2AW_2 - 
        0.2AS + 0.2W_2S + 2AW_2S - 0.2)$\\
$M \sim bernoulli $& $P(M = 1) = expit(1Z + 6W_2Z - 2W_2 - 2)$\\
$Y \sim bernoulli $& $P(Y = 1) = expit(log(1.2) + log(40)Z - log(30)M - log(1.2)
        W_2 - log(40)W_2Z)$\\
   \hline
\end{tabular}
\end{table}

\subsection{Results}
First, in Table \ref{restabcorrect}, we show results under correct specification of all models for sample sizes of N=5,000, N=500, and N=100 using DGM 1. Results using alternative DGMs were similar and are shown in the appendix. 

Table \ref{restabcorrect} shows that under correct parametric model specifications, all estimators result in consistent estimates with 95\% CI coverage close to 95\% for sample sizes of N=5,000 and N=500. Influence curve-based efficiency is close to 100\% of the efficiency bound for both the efficient TMLE and EE estimators for all sample sizes. For the transported direct effect, efficiency is about 3 times the efficiency bound for the general TMLE and EE estimators and about 4 times the efficiency bound for the stabilized IPTW estimator. For the transported indirect effect, IC- and bootstrapped-based efficiencies are similar and close to the efficiency bound for the TMLE and EE estimators. The stabilized IPTW estimator has efficiencies about 70\% larger than the efficiency bound. In the appendix, we review the theory underlying the empirical finding that the stabilized IPTW estimator is least efficient. 

Performance of all estimators degrades with a smaller sample size of N=100. For this sample size, bias slightly increases, the estimates stray from the efficiency bound, and IC-based coverage generally decreases (though bootstrapping can recover some of this lost coverage by approximating the true variance of the estimator). 

\begin{singlespacing}
\newpage
{\footnotesize
\begin{longtable}{|p{3cm}| p{1.5cm} p{1cm} p{1.7cm} p{1cm} p{1.7cm}  p{1.5cm}  p{1.5cm}  | }

\caption{Simulation results comparing estimators of the transported stochastic direct effect and transported stochastic indirect effect 
 under DGM 1 and correct model specification for various sample sizes. 1,000 simulations. Estimation methods compared include stabilized inverse probability weighting estimation (IPTW), solving the estimating equation (EE), and targeted minimum loss-based estimation (TMLE). For TMLE and EE, we compare versions of the estimators that incorporate the exclusion restrictions in our statistical model (TMLE efficient, EE efficient) and versions that do not (TMLE general, EE general). Efficiency and 95\% CI coverage are calculated separately using 1) the influence curve (IC) and 2) bootstrapping (boot). Bias and RMSE values are averages across the simulations.}
\label{restabcorrect}\\
\hline

Estimator & Bias & \multicolumn{2}{ c  }{Efficiency} & \multicolumn{2}{ c  }{95\% CI Coverage}  & RMSE & \% Out of Bounds \\  &      & \multicolumn{1}{ l  }{IC}& \multicolumn{1}{ l  }{Boot}     &  \multicolumn{1}{ l  }{IC}& \multicolumn{1}{ l  }{Boot}                 &   &  \\   
\hline 
\endfirsthead

\multicolumn{8}{c}%
{{\tablename\ \thetable{} -- continued from previous page}} \\
\hline
Estimator & Bias & \multicolumn{2}{ c  }{Efficiency} & \multicolumn{2}{ c  }{95\%CI Cov}  & RMSE &  \% Out of Bounds \\  &      & \multicolumn{1}{ l  }{IC}& \multicolumn{1}{ l  }{Boot}     &  \multicolumn{1}{ l  }{IC}& \multicolumn{1}{ l  }{Boot}                 &   &  \\   
\hline
\endhead

\hline \multicolumn{8}{|r|}{{Continued on next page}} \\ \hline
\endfoot

\hline
\endlastfoot
\multicolumn{8}{| c | }{Transported stochastic direct effect }\\ \hline
N=5000&&&&&&&\\
TMLE efficient &0.000 & 100.17 & 100.42 & 0.955 & 0.957 & 0.008 & 0 \\ 
TMLE general& 0.001 & 321.02 & 321.92 & 0.945 & 0.942 & 0.027 & 0 \\ 
EE efficient  & 0.000 & 100.21 & 100.42 & 0.955 & 0.957 & 0.008 & 0 \\ 
EE general& 0.001 & 321.27 & 321.00 & 0.946 & 0.942 & 0.027 & 0 \\ 
IPTW& 0.000 & 412.67 & 388.11 & 0.954 & 0.945 & 0.033 & 0 \\ 
\hline
N=500&&&&&&&\\  
TMLE efficient & 0.000 & 101.37 & 105.16 & 0.946 & 0.957 & 0.028 & 0 \\ 
TMLE general & -0.004 & 319.08 & 327.43 & 0.929 & 0.936 & 0.089 & 0 \\ 
EE efficient& 0.000 & 101.40 & 104.51 & 0.946 & 0.957 & 0.028 & 0 \\ 
EE general& -0.004 & 321.97 & 316.56 & 0.947 & 0.935 & 0.088 & 0 \\ 
IPTW & -0.001 & 428.43 & 419.16 & 0.951 & 0.942 & 0.110 & 0 \\ 
\hline
N=100&&&&&&&\\
TMLE efficient& 0.003 & 102.26 & 139.52 & 0.970 & 0.992 & 0.064 & 0 \\ 
TMLE general & 0.007 & 293.07 & 375.10 & 0.852 & 0.945 & 0.218 & 0 \\ 
EE efficient & 0.004 & 102.13 & $118.65$ & 0.974 & 0.989 & 0.060 & 0 \\ 
EE general& 0.005 & 316.03 & $292.87$ & 0.957 & 0.930 & 0.186 & 0 \\ 
IPTW & 0.009 & 442.80 & 456.03 & 0.909 & 0.922 & 0.267 & 0 \\ 
\hline
\multicolumn{8}{| c | }{Transported stochastic indirect effect }\\ \hline
N=5000&&&&&&&\\
TMLE efficient & 0.000 & 99.98 & 100.46 & 0.942 & 0.942 & 0.004 & 0 \\ 
TMLE general & 0.000 & 101.75 & 101.99 & 0.942 & 0.942 & 0.004 & 0 \\ 
EE efficient & 0.000 & 100.01 & 100.46 & 0.942 & 0.942 & 0.004 & 0 \\ 
EE general & 0.000 & 101.69 & 101.95 & 0.942 & 0.943 & 0.004 & 0 \\ 
IPTW& 0.000 & 166.56 & 141.45 & 0.980 & 0.957 & 0.005 & 0 \\ 
\hline
N=500&&&&&&&\\
TMLE efficient& -0.001 & 99.56 & 104.59 & 0.926 & 0.929 & 0.012 & 0 \\ 
TMLE general & -0.001 & 101.74 & 106.99 & 0.928 & 0.935 & 0.012 & 0 \\ 
EE efficient& -0.001 & 99.58 & 101.11 & 0.926 & 0.929 & 0.012 & 0 \\ 
EE general & -0.001 & 101.94 & 103.39 & 0.932 & 0.930 & 0.012 & 0 \\ 
IPTW& -0.001 & 177.68 & 156.98 & 0.948 & 0.936 & 0.018 & 0 \\
\hline
N=100&&&&&&&\\
TMLE efficient& -0.003 & 93.01 & 132.95 & 0.878 & 0.944 & 0.033 & 0 \\
TMLE general& -0.001 & 99.85 & 158.57 & 0.865 & 0.952 & 0.041 & 0 \\ 
EE efficient& -0.003 & 93.03 & 113.13 & 0.878 & 0.924 & 0.033 & 0 \\ 
EE general & -0.002 & 99.09 & $117.32$ & 0.871 & 0.915 & 0.036 & 0 \\ 
IPTW& 0.002 & 167.51 & 206.17 & 0.812 & 0.867 & 0.062 & 0 \\ \hline
\end{longtable}
}
\end{singlespacing}

Next, we show results under various model misspecifications in Table \ref{restabincorrect} for sample size N=5,000. We consider: 1) misspecification of the $Y$ model, 2) misspecification of the $Y$ and $Z$ models, 3) misspecification of the $Y$ and $M$ models, 4) misspecification of the $Y$ and $S$ models, and 5) misspecification of the $Z$, $M$, and $S$ models. We use DGM 2 for misspecification of the $Y$ model and misspecification of the $Y$ and $M$ models. We use DGM 1 for misspecification of the $Y$ and $Z$ models, misspecification of the $Y$ and $S$ models, and misspecification of the $Z,$ $M,$ and $S$ models. Full results for misspecified models under more DGMs and various sample sizes are shown in the appendix.

As derived from the robustness properties proven in the appendix, we expect the TMLE and EE estimators to be consistent if 1) the $Y$ model is correctly specified or if 2) the $Z$, $M$, and $S$ models are correctly specified. The stabilized IPTW estimator will be consistent only in the latter case: where the $Z$, $M$, and $S$ models are correctly specified. Note that we assume $A$ is randomly assigned, so we don't examine misspecification of the $A$ model. However, in cases where $A$ is nonrandom, condition 2 above would change to require correct specification of the $A$, $Z$, $M$, and $S$ models.

We see in Table \ref{restabincorrect} that, as expected, all estimators remain unbiased when the $Y$ model is misspecified. Compared to the correctly specified case (Table \ref{restabcorrect}, there is a slight reduction in efficiency, evidenced by efficiencies greater than 100\% of the efficiency bound and slight overcoverage when using the IC as opposed to bootstrapping (Table \ref{restabincorrect}). 

When the $Z, M, $ and $S$ models are misspecified, the TMLE and EE estimators also remain unbiased, as expected, but the stabilized IPTW estimator is no longer consistent, resulting in 0\% coverage of the transported SDE (Table \ref{restabincorrect}). However, in this scenario, coverage of the TMLE and EE efficient estimators is anticonservative when using the IC, evidenced by efficiencies as low as 55\% of the efficiency bound in for the transported SIE and coverage as low as 73\%. 
 This result is not unexpected---the EIC may no longer provide accurate inference under these misspecifications, because the misspecified models may bias the influence curve approximation. In this particular misspecified scenario, our influence curve approximation becomes a less variant mean-zero function of the observed data than our efficiency bound given by the true influence curve, causing poor coverage. In such instances of model misspecification, using the bootstrap can recover coverage (as long the fitting methods applied yield valid standard errors under the non-parametric bootstrap.  For linear regressions, this is the case, but for data adaptive machine learning algorithms, such might not be the case \citep{van1996weak}). 

In the remaining model misspecifications shown in Table \ref{restabincorrect}, none of the estimators are guaranteed to be unbiased. 
 In the case where the $Y$ and $M$ models are misspecified, all three estimators perform poorly in estimating both the transported direct and indirect effects. Performances of the EE 
 estimators are particularly poor in terms of bias and inefficiency. The marked inefficiency may be due, in part, to estimates lying outsides the bounds of the parameter space. Indeed, we see that nearly 100\% of the EE transported SDE estimates lie outside the parameter space. In contrast, even though the TMLE estimators solve the same influence curve, none of the TMLE estimates lie outside the parameter space, demonstrating its advantage as a substitution estimator.

\begin{singlespacing}
{\footnotesize
\begin{longtable}{|p{3cm}| p{1.5cm} p{1cm} p{1.7cm} p{1cm} p{1.7cm}  p{1.5cm}  p{1.5cm}  | }

\caption{Simulation results comparing estimators of transported stochastic direct effect and transported stochastic indirect effect under various model misspecifications for sample size N=5,000. 1,000 simulations. Estimation methods compared include stabilized inverse probability weighting estimation (IPTW), solving the estimating equation (EE), and targeted minimum loss-based estimation (TMLE).  For TMLE and EE, we compare versions of the estimators that incorporate the exclusion restrictions in our statistical model (TMLE efficient, EE efficient) and versions that do not (TMLE general, EE general).
Bias and RMSE values are averages across the simulations.
Efficiency and 95\% CI coverage are calculated separately using 1) the influence curve (IC) and 2) bootstrapping (boot). The DGM used for each misspecification is noted below.}
\label{restabincorrect}\\
\hline

Estimator & Bias & \multicolumn{2}{ c  }{Efficiency} & \multicolumn{2}{ c  }{95\% CI Coverage}  & RMSE & \% Out of Bounds \\  &      & \multicolumn{1}{ l  }{IC}& \multicolumn{1}{ l  }{Boot}     &  \multicolumn{1}{ l  }{IC}& \multicolumn{1}{ l  }{Boot}                 &   &  \\   
\hline 
\endfirsthead

\multicolumn{8}{c}%
{{ \tablename\ \thetable{} -- continued from previous page}} \\
\hline
Estimator & Bias & \multicolumn{2}{ c  }{Efficiency} & \multicolumn{2}{ c  }{95\% CI Coverage}  & RMSE &  \% Out of Bounds \\  &      & \multicolumn{1}{ l  }{IC}& \multicolumn{1}{ l  }{Boot}     &  \multicolumn{1}{ l  }{IC}& \multicolumn{1}{ l  }{Boot}                 &   &  \\   
\hline
\endhead

\hline \multicolumn{8}{|r|}{{Continued on next page}} \\ \hline
\endfoot

\hline
\endlastfoot

\multicolumn{8}{| c | }{Y model misspecified, DGM 2} \\ \hline
\multicolumn{8}{| c | }{Transported stochastic direct effect } \\ \hline
TMLE efficient & 0.000 & 191.06 & 118.82 & 0.999 & 0.965 & 0.014 & 0 \\ 
TMLE general& 0.000 & 354.81 & 288.79 & 0.984 & 0.954 & 0.035 & 0 \\ 
EE efficient & 0.000 & 240.26 & 126.67 & 0.999 & 0.968 & 0.015 & 0 \\ 
EE general & 0.000 & 374.95 & 291.49 & 0.992 & 0.957 & 0.035 & 0 \\ 
IPTW & 0.000 & 300.52 & 276.72 & 0.973 & 0.957 & 0.034 & 0 \\ 
\hline
\multicolumn{8}{| c | }{Transported stochastic indirect effect} \\ \hline
TMLE efficient & 0.000 & 186.27 & 102.20 & 0.995 & 0.914 & 0.002 & 0 \\ 
TMLE general & 0.000 & 239.35 & 186.41 & 0.986 & 0.942 & 0.004 & 0 \\ 
EE efficient& 0.000 & 237.87 & 103.75 & 0.997 & 0.919 & 0.002 & 0 \\ 
EE general & 0.000 & 302.49 & 216.48 & 0.991 & 0.945 & 0.004 & 0 \\ 
IPTW & 0.000 & 223.45 & 174.48 & 0.988 & 0.941 & 0.003 & 0 \\ 
\hline
\multicolumn{8}{| c | }{Y and Z models misspecified, DGM 1} \\ \hline
\multicolumn{8}{| c | }{Transported stochastic direct effect} \\ \hline 
TMLE efficient& 0.032 & 75.49 & 77.35 & 0.000 & 0.000 & 0.032 & 0 \\ 
TMLE general& 0.188 & 973.02 & 463.63 & 0.223 & 0.003 & 0.192 & 0 \\ 
EE efficient& 0.033 & 80.22 & 81.92 & 0.000 & 0.000 & 0.034 & 0 \\ 
EE general& 0.545 & 1,112.09 & 1,008.79 & 0.000 & 0.000& 0.552 & 0 \\ 
IPTW & 0.108 & 700.28 & 456.29 & 0.574 & 0.204 & 0.115 & 0 \\ 
\hline
\multicolumn{8}{| c | }{Transported stochastic indirect effect} \\ \hline
TMLE efficient & -0.014 & 192.60 & 92.53 & 0.380 & 0.004 & 0.014 & 0 \\ 
TMLE general & -0.040 & 492.45 & 143.03 & 0.185 & 0.000 & 0.040 & 0 \\ 
EE efficient& 0.018 & 307.60 & 121.67 & 0.767 & 0.029 & 0.019 & 0 \\ 
EE general& 0.066 & 911.65 & 686.60 & 0.447 & 0.231 & 0.072 & 0 \\ 
IPTW  & -0.031 & 439.30 & 160.87 & 0.430 & 0.004 & 0.031 & 0 \\ 
\hline
\multicolumn{8}{| c | }{Y and M models misspecified, DGM 2} \\ \hline
\multicolumn{8}{| c | }{Transported stochastic direct effect } \\ \hline
TMLE efficient & -0.272 & 280.08 & 149.17 & 0.000 & 0.000 & 0.273 & 0 \\ 
TMLE general& -0.272 & 698.69 & 321.77 & 0.147 & 0.001 & 0.276 & 0 \\ 
EE efficient& -1.110 & 852.67 & 735.38 & 0.000 & 0.000 & 1.118 & 99.2 \\ 
EE general & -1.111 & 1,093.33 & 877.83 & 0.000 & 0.000 & 1.121 & 98.4 \\ 
IPTW & -0.236 & 589.90 & 284.02 & 0.090 & 0.000 & 0.240 & 0 \\ 
\hline
\multicolumn{8}{| c | }{Transported stochastic indirect effect} \\ \hline
TMLE efficient& 0.104 & 249.26 & 173.89 & 0.000 & 0.000 & 0.106 & 0 \\ 
TMLE general & 0.103 & 370.51 & 279.70 & 0.036 & 0.016 & 0.106 & 0 \\ 
EE efficient & 0.135 & 336.39 & 227.03 & 0.000 & 0.000 & 0.137 & 0 \\ 
EE general& 0.137 & 535.91 & 413.33 & 0.054 & 0.016 & 0.141 & 0 \\ 
IPTW & 0.061 & 279.01 & 200.78 & 0.197 & 0.090 & 0.063 & 0 \\ 
\hline
\multicolumn{8}{| c | }{Y and S models misspecified, DGM 1} \\ \hline
\multicolumn{8}{| c | }{Transported stochastic direct effect  } \\ \hline
TMLE efficient & -0.164 & 285.91 & 168.14 & 0.000 & 0.000 & 0.165 & 0 \\ 
TMLE general& -0.164 & 511.22 & 399.63 & 0.019 & 0.010 & 0.167 & 0 \\ 
EE efficient & -0.220 & 415.64 & 191.91 & 0.000 & 0.000 & 0.220 & 0 \\
EE general & -0.218 & 699.90 & 586.82 & 0.029 & 0.023 & 0.224 & 0 \\ 
IPTW& -0.067 & 404.82 & 322.51 & 0.467 & 0.319 & 0.073 & 0 \\ 
\hline
\multicolumn{8}{| c | }{Transported stochastic indirect effect } \\ \hline
TMLE efficient & 0.222 & 507.95 & 186.55 & 0.000 & 0.000 & 0.223 & 0 \\ 
TMLE general& 0.222 & 600.12 & 343.91 & 0.000 & 0.000& 0.222 & 0 \\ 
EE efficient & 0.278 & 1,033.84 & 216.14 & 0.000 & 0.000 & 0.279 & 0 \\ 
EE general & 0.278 & 1,201.77 & 675.26 & 0.000 & 0.000 & 0.279 & 0 \\ 
IPTW & 0.278 & 520.31 & 337.87 & 0.000 & 0.000 & 0.278 & 0 \\ 
\hline
\multicolumn{8}{| c | }{Z, M, and S models misspecified, DGM 1} \\ \hline
\multicolumn{8}{| c | }{Transported stochastic direct effect  } \\ \hline
TMLE efficient& 0.000 & 98.78 & 100.01 & 0.956 & 0.960 & 0.008 & 0 \\ 
TMLE general & 0.001 & 228.37 & 231.43 & 0.939 & 0.941 & 0.020 & 0 \\ 
EE efficient& 0.000 & 98.80 & 100.03 & 0.956 & 0.960 & 0.008 & 0 \\ 
EE general& 0.001 & 228.50 & 203.24 & 0.966 & 0.942 & 0.018 & 0 \\ 
IPTW & 0.254 & 396.11 & 406.66 & 0.000 & 0.000 & 0.256 & 0 \\ 
\hline
\multicolumn{8}{| c | }{Transported stochastic indirect effect } \\ \hline
TMLE efficient& 0.001 & 54.95 & 100.51 & 0.727 & 0.938 & 0.010 & 0 \\ 
TMLE general& 0.001 & 129.68 & 136.04 & 0.879 & 0.920 & 0.014 & 0 \\ 
EE efficient & 0.000 & 54.95 & 100.46 & 0.732 & 0.939 & 0.010 & 0 \\ 
EE general & 0.001 & 130.08 & 137.03 & 0.891 & 0.930 & 0.014 & 0 \\ 
IPTW & 0.014 & 301.92 & 123.87 & 0.976 & 0.716 & 0.018 & 0 \\ 
\hline

\end{longtable}
}
\end{singlespacing}


\section{Conclusion}
In this paper, we defined and identified parameters that transport stochastic direct and indirect mediating effects from a source population ($S=1$) to a new, target population ($S=0$). Identification of such parameters rely on the typical sequential randomization and positivity assumptions of other stochastic mediation effects \citep{rudolph2017robust,zheng2017longitudinal,vanderweele2016mediation} as well as a common outcome model assumption, described previously for transport estimators \citep{rudolph2016robust}, which can be tested nonparametrically \citep{luedtke2015omnibus}. Such parameters enable the prediction of mediating effects in new populations based on data about the mediation mechanism in a source population and the differing distributions of compositional characteristics between the two populations. Thus, transport SDE and SIE parameters contribute to understanding the how and why interventions may work differently and/or have differing effects when applied to new populations. 

We proposed three estimation approaches for such effects: a stabilized IPTW estimator, an estimating equation approach, and a TMLE approach. For the EE and TMLE approaches, we describe a version of each estimator that is efficient under a statistical model with exclusion restrictions such that $A$ does not directly affect $M$ or $Y$, and another version that is efficient under a statistical model without those restrictions such that $A$ can directly affect $M$ and $Y$. The EE and TMLE estimators solve the EIC for a particular statistical model, which results in double robustness, meaning that they are unbiased if one either consistently estimates the $Y$ model or $Z, M, S$ models. The TMLE estimator has the additional advantage of staying within the bounds of the parameter space by virtue of being a substitution estimator. We demonstrated the finite sample advantage of staying within the parameter bounds in a simulation study; in a particularly challenging scenario, nearly all of the EE estimates were outside of the parameter space but the TMLE estimates stayed within bounds. The simulation study also demonstrated that, as expected, the stabilized IPTW estimator is less efficient than the EE and TMLE estimators under correct model specification and in many scenarios where models are misspecified. 

We also saw empirical evidence that, even in cases where robustness properties guarantee estimator consistency, the EIC may no longer provide accurate inference when models are misspecified. If parametric models are used, as in our simulation, the bootstrap can be used to recover appropriate coverage in such scenarios. However, the bootstrap is not an appropriate strategy if data-adaptive methods are used in model fitting. An alternative approach is to target the nuisance parameters of the IC (e.g, $g_M$, $g_Z$, $g_S$, etc.) in addition to the parameter of interest in a TMLE \citep{benkeser2017doubly}. Applying such extra targeting the to TMLE estimators we describe here is an area for future work. 

To facilitate use of these methods, we include commented R code for implementing them in the appendix. 

The estimators we propose are limited in that they consider a stochastic intervention on mediator, $M$, that is assumed known and estimated from observed data. However, we plan to extend them to a true, unknown stochastic intervention in the future. Another limitation is that the parameters are only identified if one assumes a common outcome model across the source and target populations. There will be some research questions for which it is not possible to establish evidence for or against this assumption, as in questions about predicting a long-term outcome in a new population. However, when the research question instead focuses on establishing the extent to which mechanisms are shared across populations, and the full set of data $O=(S,W,A,Z,M,Y)$ is observed for both populations, one can empirically test whether there is evidence against such a shared outcome model \citep{luedtke2015omnibus}. 

In the main text, we focused on transporting mediation estimates where an instrument, $A$, was statically intervened on and mediator $M$ was stochastically intervened on. Moreover, we were primarily concerned with a statistical model that imposed instrumental variable assumptions such as the exclusion restriction assumption. However, we describe how each estimator can be easily modified to accommodate statistical models that do not impose instrumental variable assumptions, allowing for a direct effect of $A$ on $M$ and of $A$ on $Y$. Extending our proposed estimators for data generating mechanisms that do not include an intermediate variable, $Z$ is straightforward. Thus, our transport mediation estimators can be applied to a wide-range of common data generating mechanisms. 

\bibliographystyle{rss}
\bibliography{lib}

\begin{thebibliography}{27}
\expandafter\ifx\csname natexlab\endcsname\relax\def\natexlab#1{#1}\fi
\expandafter\ifx\csname url\endcsname\relax
  \def\url#1{\texttt{#1}}\fi
\expandafter\ifx\csname urlprefix\endcsname\relax\def\urlprefix{URL}\fi

\bibitem[{Angrist et~al.(1996)Angrist, Imbens and
  Rubin}]{angrist1996identification}
Angrist, J.~D., Imbens, G.~W. and Rubin, D.~B. (1996) Identification of causal
  effects using instrumental variables.
\newblock \textit{Journal of the American statistical Association},
  \textbf{91}, 444--455.

\bibitem[{Bareinboim and Pearl()}]{bareinboim2013general}
Bareinboim, E. and Pearl, J. () A general algorithm for deciding
  transportability of experimental results.
\newblock \textit{Journal of causal Inference}, \textbf{1}, 107--134.

\bibitem[{Benkeser et~al.(2017)Benkeser, Carone, Laan and
  Gilbert}]{benkeser2017doubly}
Benkeser, D., Carone, M., Laan, M. V.~D. and Gilbert, P. (2017) Doubly robust
  nonparametric inference on the average treatment effect.
\newblock \textit{Biometrika}, \textbf{104}, 863--880.

\bibitem[{Cole and Stuart(2010)}]{cole2010generalizing}
Cole, S.~R. and Stuart, E.~A. (2010) Generalizing evidence from randomized
  clinical trials to target populations: The actg 320 trial.
\newblock \textit{American journal of epidemiology}, \textbf{172}, 107--115.

\bibitem[{Frangakis(2009)}]{frangakis2009calibration}
Frangakis, C. (2009) The calibration of treatment effects from clinical trials
  to target populations.
\newblock \textit{Clinical trials (London, England)}, \textbf{6}, 136.

\bibitem[{Kling et~al.(2007)Kling, Liebman and Katz}]{kling2007experimental}
Kling, J.~R., Liebman, J.~B. and Katz, L.~F. (2007) Experimental analysis of
  neighborhood effects.
\newblock \textit{Econometrica}, \textbf{75}, 83--119.

\bibitem[{van~der Laan(2016)}]{Laan:2015aa}
van~der Laan, M. (2016) A generally efficient targeted minimum loss based
  estimator.
\newblock \textit{U.C. Berkeley Division of Biostatistics Working Paper
  Series}, \textbf{343}.
\newblock \urlprefix\url{http://biostats.bepress.com/ucbbiostat/paper343}.

\bibitem[{van~der Laan and Rose(2011)}]{Laan:2011aa}
van~der Laan, M. and Rose, S. (2011) \textit{Targeted Learning}.
\newblock New York: Springer.

\bibitem[{van~der Laan and Rubin(2006)}]{Laan:2006aa}
van~der Laan, M. and Rubin, D. (2006) Targeted maximum likelihood learning.
\newblock \textit{U.C. Berkeley Division of Biostatistics Working Paper
  Series}.
\newblock \urlprefix\url{http://biostats.bepress.com/ucbbiostat/paper213}.

\bibitem[{van~der Laan and Gruber(2012)}]{van2012targeted}
van~der Laan, M.~J. and Gruber, S. (2012) Targeted minimum loss based
  estimation of causal effects of multiple time point interventions.
\newblock \textit{The international journal of biostatistics}, \textbf{8}.

\bibitem[{Levy(2019)}]{levy2019tutorial}
Levy, J. (2019) Tutorial: Deriving the efficient influence curve for large
  models.
\newblock \textit{arXiv preprint arXiv:1903.01706}.

\bibitem[{Luedtke et~al.(2015)Luedtke, Carone and van~der
  Laan}]{luedtke2015omnibus}
Luedtke, A.~R., Carone, M. and van~der Laan, M.~J. (2015) An omnibus
  nonparametric test of equality in distribution for unknown functions.
\newblock \textit{arXiv preprint arXiv:1510.04195}.

\bibitem[{Miettinen(1972)}]{miettinen1972standardization}
Miettinen, O.~S. (1972) Standardization of risk ratios.
\newblock \textit{American Journal of Epidemiology}, \textbf{96}, 383--388.

\bibitem[{Neyman(1923)}]{neyman1923application}
Neyman, J.~S. (1923) On the application of probability theory to agricultural
  experiments. essay on principles. section 9.(tlanslated and edited by dm
  dabrowska and tp speed, statistical science (1990), 5, 465-480).
\newblock \textit{Annals of Agricultural Sciences}, \textbf{10}, 1--51.

\bibitem[{Pearl(2009)}]{pearl2009causality}
Pearl, J. (2009) \textit{Causality}.
\newblock Cambridge university press.

\bibitem[{Pearl and Bareinboim(2014)}]{pearl2014external}
Pearl, J. and Bareinboim, E. (2014) External validity: From do-calculus to
  transportability across populations.
\newblock \textit{Statistical Science}, 579--595.

\bibitem[{Pearl et~al.(2009)}]{pearl2009causal}
Pearl, J. et~al. (2009) Causal inference in statistics: An overview.
\newblock \textit{Statistics surveys}, \textbf{3}, 96--146.

\bibitem[{Rubin(1974)}]{rubin1974estimating}
Rubin, D.~B. (1974) Estimating causal effects of treatments in randomized and
  nonrandomized studies.
\newblock \textit{Journal of educational Psychology}, \textbf{66}, 688.

\bibitem[{Rudolph et~al.(2018{\natexlab{a}})Rudolph, Goin, Paksarian, Crowder,
  Merikangas and Stuart}]{rudolph2018causal}
Rudolph, K.~E., Goin, D.~E., Paksarian, D., Crowder, R., Merikangas, K.~R. and
  Stuart, E.~A. (2018{\natexlab{a}}) Causal mediation analysis with
  observational data: Considerations and illustration examining mechanisms
  linking neighborhood poverty to adolescent substance use.
\newblock \textit{American journal of epidemiology}.

\bibitem[{Rudolph and van~der Laan(2017)}]{rudolph2016robust}
Rudolph, K.~E. and van~der Laan, M.~J. (2017) Robust estimation of
  encouragement-design intervention effects transported across sites.
\newblock \textit{Journal of the Royal Statistical Society Series B Statistical
  Methodology}, \textbf{79}, 1509--1525.

\bibitem[{Rudolph et~al.(2018{\natexlab{b}})Rudolph, Schmidt, Crowder, Galin,
  Glymour, Ahern and Osypuk}]{rudolph2018composition}
Rudolph, K.~E., Schmidt, N.~M., Crowder, R., Galin, J., Glymour, M.~M., Ahern,
  J. and Osypuk, T.~L. (2018{\natexlab{b}}) Composition or context: using
  transportability to understand drivers of site differences in a large-scale
  housing experiment.
\newblock \textit{Epidemiology}, \textbf{29}, 199--206.

\bibitem[{Rudolph et~al.(2017)Rudolph, Sofrygin, Zheng and van~der
  Laan}]{rudolph2017robust}
Rudolph, K.~E., Sofrygin, O., Zheng, W. and van~der Laan, M.~J. (2017) Robust
  and flexible estimation of data-dependent stochastic mediation effects: a
  proposed method and example in a randomized trial setting.
\newblock \textit{Epidemiologic Methods}, In Press.

\bibitem[{Stuart et~al.(2011)Stuart, Cole, Bradshaw and Leaf}]{stuart2011use}
Stuart, E.~A., Cole, S.~R., Bradshaw, C.~P. and Leaf, P.~J. (2011) The use of
  propensity scores to assess the generalizability of results from randomized
  trials.
\newblock \textit{Journal of the Royal Statistical Society: Series A
  (Statistics in Society)}, \textbf{174}, 369--386.

\bibitem[{Van Der~Vaart and Wellner(1996)}]{van1996weak}
Van Der~Vaart, A.~W. and Wellner, J.~A. (1996) Weak convergence.
\newblock In \textit{Weak convergence and empirical processes}, 16--28.
  Springer.

\bibitem[{VanderWeele and Tchetgen~Tchetgen(2017)}]{vanderweele2016mediation}
VanderWeele, T.~J. and Tchetgen~Tchetgen, E.~J. (2017) Mediation analysis with
  time varying exposures and mediators.
\newblock \textit{Journal of the Royal Statistical Society: Series B
  (Statistical Methodology)}, \textbf{79}, 917--938.

\bibitem[{Wright(1921)}]{wright1921correlation}
Wright, S. (1921) Correlation and causation.
\newblock \textit{Journal of agricultural research}, \textbf{20}, 557--585.

\bibitem[{Zheng and van~der Laan(2017)}]{zheng2017longitudinal}
Zheng, W. and van~der Laan, M. (2017) Longitudinal mediation analysis with
  time-varying mediators and exposures, with application to survival outcomes.
\newblock \textit{Journal of Causal Inference}.

\end{thebibliography}

\newpage
\appendix

\section{Identifiability}
To identify the stochastic direct effect and stochastic indirect effect we will need to impose additional assumptions on $\mathcal{M}^F$ and $\mathcal{M}$, listed below.  

\begin{enumerate}
\item Positivity: For all $S$ and $W$ we need a positive probability of
assigning any level of $A$. For all combinations of $S, W,$ and $A=a$, we have a positive probability of any level of $Z.$ For
$S=1$ and all combinations of $Z$ and $W$ we need a positive probability
of any level of the mediator, $M$.  

\item Common outcome model across sites: $\mathbb{E}\left[Y\mid M,Z,W,S=1\right]=\mathbb{E}\left[Y\mid M,Z,W,S=0\right]$. The null hypothesis of a common outcome model may be tested nonparametrically \citep{luedtke2015omnibus}.

\item Sequential Randomization: $Y_{am}\perp A\mid W,S$ and $Y_{am}\perp M\mid W,Z,S$.  
This is akin to a two-time point longitudinal intervention where at the first time point, we statically intervene to set the treatment, $A=a$, and at the second time point, we stochastically intervene on the mediator, $M$.  
\end{enumerate}

\begin{thm}
\begin{align*}
& \Psi(P) = \Psi^F(P_{UX}) = \mathbb{E}\biggr[\mathbb{E}\biggr[\mathbb{E}_{\hat{g}_{M \mid a^*,W,s}}\biggr[\mathbb{E}\biggr[Y | M, Z,W, S=1 \biggr] | W,Z\biggr] | W,a,S=0\biggr] | S=0\biggr] \\
&= \mathbb{E} \biggr[ \mathbb{E}\biggr[\sum_m \biggr[  \mathbb{E}Y\hat{g}_{M\mid a^*,W,s}(m \mid W) \mid M=m, Z, A=a, W, S =1 \biggr] \mid A=a, W,S \biggr] \mid S = 0 \biggr]
\end{align*}
\end{thm}
Proof:

\begin{align*}
\Psi^F(P_{UX})&=\mathbb{E}\biggr[Y_{a,\hat{g}_{M\mid a^{*},W,s}}\mid S=0\biggr]\\
& \overset{\text{tower law}}{=}  \mathbb{E} \biggr[ \mathbb{E} \biggr[Y_{a,\hat{g}_{M\mid a^*,W,s}} \mid Z,A,W \biggr] \mid S = 0 \biggr]\\
& \overset{\text{assumption 2}}{=}  \mathbb{E}\biggr[\sum_m \biggr[  \mathbb{E}Y_{am} \hat{g}_{M\mid a^*,W,s}(m \mid W) \mid M=m,Z,A, W, S =1 \biggr]  \mid S = 0 \biggr]\\
& \overset{\text{tower law}}{=}  \mathbb{E} \biggr[ \mathbb{E}\biggr[\sum_m \biggr[  \mathbb{E}Y_{am} \hat{g}_{M\mid a^*,W,s}(m \mid W) \mid M=m,Z,A, W, S =1 \biggr]\\ 
& \hspace{2cm} \mid A, W,S=0 \biggr] \mid S = 0 \biggr]\\
& \overset{\text{assumption 3}}{=}  \mathbb{E} \biggr[ \mathbb{E}\biggr[\sum_m \biggr[  \mathbb{E}Y \hat{g}_{M\mid a^*,W,s}(m \mid W) \mid M=m, Z,A=a, W, S =1 \biggr]\\
& \hspace{2cm} \mid A=a, W,S=0 \biggr] \mid S = 0 \biggr]
\end{align*}

The conditional expectations are well-defined due to assumption 1 above.

\section{Estimator derivations}
\subsection{Stabilized inverse probability of treatment weighted estimator}
This estimator solves the estimating equation $P_n \hat{H}_n(Y- \Psi_n) =0$, where $\hat{H}_n(M,Z,A,W,S)$ is an estimate of the true $H$ given by
\begin{footnotesize}
\begin{equation}
H(M,Z,A,W,S)= \frac{\hat{g}_{M\mid a^{*},W,s}\left(M\mid W\right)p_{Z}\left(Z\mid A,W,S=0\right)p_{S\mid W}\left(S=0\mid W\right)I(S=1,A=a)}{p_{M}\left(M\mid Z,W,S\right)p_{Z}\left(Z\mid A,W,S\right)p_{A}\left(A\mid W,S\right)p_{S \mid W}\left(S\mid W\right)P_{S}(S=0)}.
\end{equation}
\end{footnotesize}

To solve the estimating equation, we find an initial estimate, $\Psi_n^0 = P_n \hat{H}_n Y$, i.e., the standard unstabilized IPTW estimator.  
$$\Psi_n^1 - \Psi_n^0 = -\left(\frac{d}{d\Psi}P_n \hat{H}_n (Y - \Psi_n^0)\right)^{-1}P_n \hat{H}_n(Y - \Psi_n^0),$$ where our stablilized estimator is then $\Psi_n^{IPTW} = \Psi_n^1 = P_n \hat{H}_n Y/P_n \hat{H}_n$.  

\subsubsection{Efficiency}
\begin{thm}
Assume
\begin{enumerate}
\item
A1: $\Vert H - \hat{H}_n\Vert_{L^2(P_0)}=o_P(n^{-0.5})$ 
\item
A2: $\hat{H}_n$ and $\hat{D}_n$ are in a $P_0$ Donsker class.  
\end{enumerate}

Let $IC_{IPTW}$ be the true influence curve or $H(Y - \Psi(P_0))/P_0 H = H(Y - \Psi(P_0))$, noting $P_0 H = 1$ for the true $H$.  Then $\sqrt{n}(\Psi^{IPTW}_{n} - \Psi_0(P))$ converges in distribution to $\sqrt{n}$ times the empirical mean of the true influence curve.  In this case, our standard error times $\sqrt{n}$ will converge to the standard deviation of $IC_{IPTW}$.  
\end{thm}
\label{iptwR2thm}
Proof:

\begin{align*}
\Psi_n^{IPTW} - \Psi(P_0) &= \Psi_n^0 + P_n \hat{D}_n - \Psi(P_0) \\
& = \Psi_n^0 + (P_n -P_0)\hat{D}_n + P_0 \hat{D}_n - \Psi(P_0)
\end{align*}

By A2 $\sqrt{n}(P_n -P_0)\hat{D}_n$ converges in distribution to a normal distribution of mean 0 and variance equal to the true variance of $IC_{IPTW}$.  Thus, we must show $\Psi_n^0 +  P_0 \hat{D}_n - \Psi(P_0) = o_P(1/\sqrt{n})$ under A1.  

\begin{footnotesize}
\begin{align}
\Psi_n^0 +  P_0 \hat{D}_n - \Psi(P_0) &= P_n \hat{H}_nY - P_0 HY +\frac{P_0 \hat{H}_n Y}{P_n \hat{H}_n} - \frac{P_n \hat{H}_n P_n \hat{H}_nY}{P_n \hat{H}_n} \nonumber \\
&=  \frac{(P_n - P_0) \hat{H}_n \Psi_n^0}{P_n \hat{H}_n}  + P_0(\hat{H}_n - H)Y + \frac{P_0 \hat{H}_n Y}{P_n \hat{H}_n}\left((P_0 - P_n)\hat{H}_n + P_0(H - \hat{H}_n)\right)\nonumber \\
&= P_0(\hat{H}_n - H)Y+ \frac{(P_0 - P_n) \hat{H}_n Y (P_0 - P_n)\hat{H}_n}{P_n \hat{H}_n} + \frac{P_0 \hat{H}_n Y}{P_n \hat{H}_n}P_0(H - \hat{H}_n) \nonumber \\
&= P_0(\hat{H}_n - H)Y+ \frac{(P_0 - P_n) \hat{H}_n Y ((P_0 - P_n)(\hat{H}_n-H)+H)}{P_n \hat{H}_n} + \frac{P_0 \hat{H}_n Y}{P_n \hat{H}_n}P_0(H - \hat{H}_n) \label{iptwR2}
\end{align}
\end{footnotesize}

We can notice in Equation \ref{iptwR2} that the first term is bounded above by $C\Vert \hat{H}_n - H \Vert_{L^2(P_0)}$ for some constant $C$.  By A2, the second term is $o_P(1/\sqrt{n})$.  By A1 the third term is also $o_P(1/\sqrt{n})$.  

\begin{thm}
The IPTW estimator's limiting variance under the assumptions of theorem \ref{iptwR2thm} is larger than that of the TMLE or EE under well-specified models, if the model is non-parametric.  If we use the TMLE and EE incorporating the efficient influence curve for the restricted model, then these are even more efficient.  

\end{thm}
Proof:\\
Let's assume the unrestricted model, $\mathcal{M}^{np}$.  We project the IPTW influence curve onto the tangent space subspace, $T_{Y\times S}$ constisting of $\{\gamma(O) \mid \mathbb{E} [\gamma(O) \mid M, Z, A, W,S] = 0$ via the formula 

$$\prod \left( IC_{IPTW} \Vert T_{Y\times S} \right) = IC_{IPTW}(O)  - \mathbb{E}\left[ IC_{IPTW} (O) \mid M,Z,A,W,S \right] = D_Y^*(O)$$ 

Likewise we have $\prod \left(IC_{IPTW}(O) \Vert T_{Z} \right) = \mathbb{E}\left[ IC_{IPTW}(O) \mid Z,A,W,S\right]  - \mathbb{E}\left[ IC_{IPTW}(O)  \mid A,W,S \right] = D_Z^*(O)$, where $T_{Z}$ is the tangent space consisting of $\{\gamma(O) \mid \mathbb{E} [\gamma(O) \mid A, W,S] = 0$.  The empirical mean of $\prod \left(IC_{IPTW}(O) \Vert T_{S} \right) = \mathbb{E}\left[ IC_{IPTW}(O) \mid W\right]  - \mathbb{E}\left[ IC_{IPTW}(O)  \mid W \right]$ = the empirical mean of $D_W^*(O)$, the $T_S$ component of the efficient influence curve for $\mathcal{M}^{np}$. We have non-trivial projections of $IC_{IPTW}$ onto $T_M$ and $T_A$.  Since the asymptotic variance of $IC_{IPTW}(O)$ is a sum of the variances of its orthogonal components, it must exceed that of the EE and TMLE estimators, and the proof is complete.    

\subsubsection{Robustness}
The IPTW estimator requires all models for $M$, $Z$, $A$ and $S$ to be estimated at parametric rates in order to be consistent.

\subsection{Estimating equation estimator}
\subsubsection{Derivation of efficient influence curve for model without exclusion restrictions}
\label{eicunrestrict}
First we derive the efficient influence curve (EIC) for the less restricted case where $A$ may directly affect $M$ and $Y$. For more information on deriving an efficient influence curve in general, we refer the reader to a tutorial by Levy, 2019 \citep{levy2019tutorial}. We will also allow our treatment mechanism not to be known, which will have no effect on the derivation but means that the EIC is applicable in situations when treatment is not randomly assigned. 
\label{ICSDE}
\begin{thm}
Consider a non-parametric model or semiparametric model with one or both the treatment and mediator mechanisms known (mechanisms for $A$ and $M$).  Consider the parameter defined by 
\[
\Psi(P) = \mathbb{E} \biggr[ \mathbb{E}\biggr[\sum_m \biggr[  \mathbb{E}Y\hat{g}_{M\mid a^*,W,s}(m \mid W) \mid M=m, W,Z,A=a, S =1 \biggr] \mid A=a, W,S \biggr] \mid S = 0 \biggr]
\]
where the expectations are taken with respect to $P$.  Then the efficient influence curve is given by
\[
D^{*}(P)(O)=D_{Y}^{*}(P)(O)+D_{Z}^{*}(P)(O)+D_{W}^{*}(P)(O)
\]
where 
\begin{footnotesize}
\begin{align*}
D_{Y}^{*}(P)(O)&=\left(Y-\mathbb{E}\left[Y\mid M,Z,A,W\right]\right)\times\\
&\frac{\hat{g}_{M\mid a^{*},W,s}\left(M\mid W\right)p_{Z}\left(Z\mid A,W,S=0\right)p_{S \mid W}\left(S=0\mid W\right)I(S=1,A=a)}{g_{M}\left(M\mid Z,A,W,S\right)p_{Z}\left(Z\mid A,W,S\right)p_{A}\left(A\mid W,S\right)p_{S \mid W}\left(S\mid W\right)P_{S}(S=0)}\\
D_{Z}^{*}(P)(O)&=\left(\bar{Q}_{M}(Z,A,W,S)-\bar{Q}_{Z}(a, W, S)\right)\frac{I(S=0,A=a)}{p_{A}\left(a\mid W,S\right)p_{S}(S=0)}\\
D_{W}^{*}(P)(O)&=\left(\bar{Q}_{Z}(a, W, S)-\Psi(P)\right)\frac{I(S=0)}{p_{S}(S=0)}
\end{align*}
\end{footnotesize}
\end{thm}
Proof: 
The density factorizes as follows:
\begin{footnotesize}
$p(O)=p_{Y\times S}(Y \times \gamma \mid M,Z,A,W, S)p_{M}(M\mid Z,A,W,S)p_{Z}(Z\mid A,Z,W,S)p_{A}(A\mid W,S)p_{W \mid S}(W\mid S)p_{S}(S)$
\end{footnotesize}

This means 
\begin{scriptsize}
$p_{\epsilon}(O)=p_{Y\times S,\epsilon}(Y \times S \mid M,Z,A,W,S)p_{M,\epsilon}(M\mid Z,A,W,S)p_{Z,\epsilon}(Z\mid A,W,S)p_{A,\epsilon}(A\mid W,S)p_{W\mid S,\epsilon}(W\mid S)p_{S,\epsilon}(S)$
\end{scriptsize}

\begin{scriptsize}
\begin{align}
\frac{d}{d\epsilon}\left(p_{Y,\epsilon}(Y\times S\mid M,Z,A,W,S)\right)\biggr\vert_{\epsilon=0} & = \left(\gamma(O)-\mathbb{E}\left[\gamma(O)\mid M,Z,A,W,S\right]\right)p_{Y}(Y \times S\mid M,Z,A,W,S)  \label{eq:3.51}\\
\frac{d}{d\epsilon}\left(p_{Z,\epsilon}(Z\mid A,W,S)\right)\biggr\vert_{\epsilon=0} & = \left(\mathbb{E}\left[\gamma(O)\mid Z,A,W,S\right]-\mathbb{E}\left[\gamma(O)\mid A,W,S\right]\right)p_{Z}(Z\mid A,W,S) \label{eq:3.52} \\
\frac{d}{d\epsilon}\left(p_{W \mid S,\epsilon}(W\mid S)\right)\biggr\vert_{\epsilon=0} & = \left(\mathbb{E}\left[\gamma(O)\mid W,S\right]-\mathbb{E}\left[\gamma(O)\mid S\right]\right)p_{W \mid S}(W\mid S) \label{eq:3.53}\
\end{align}
\end{scriptsize}

Our estimand, which identifies our parameter of interest (see Section 3 of the main text) is given by 
\begin{scriptsize}
\[
\Psi(P)=\int yp_{Y}(y\mid m,z,a,w,s=1)\hat{g}_{M\mid a^{*},W,s}\left(m\mid a^{*},w\right)p_{Z}\left(z\mid a,w,s=0\right)p_{W \mid S}\left(w\mid s=0\right)dv(y,m,z,w)
\]
\end{scriptsize}

We then take the pathwise derivative for a path along score,
$\gamma$.  We note that this derivative is unaffected by knowledge of the treatment mechanism, $E[A \mid S,W]$, or the mediator mechansim, $E[M \mid Z,A,W,S]$, due to the estimand not depending on these models as well as the fact that scores, $\gamma_A$ and $\gamma_M$ are orthogonal (have 0 covariance) to $\gamma_Y, \gamma_Z, \gamma_W)$ in the Hilbert Space $L^2(P)$.  This is why for a semi-parametric model where the treatment mechanism is known, the efficient influence curve will be the same as that for $\mathcal{M}^{np}$. 

\begin{scriptsize}
\begin{align}
\frac{d}{d\epsilon}\Psi(P_{\epsilon})\biggr\vert_{\epsilon=0} & = \frac{d}{d\epsilon}\int yp_{Y,\epsilon}(y\mid m,z,a,w,s=1)\hat{g}_{M\mid a^{*},W,s}\left(m\mid w\right)p_{Z,\epsilon}\left(z\mid a,w,s=0\right)p_{W \mid S,\epsilon}\left(w\mid s=0\right)dv(y,m,z,w)\biggr\vert_{\epsilon=0}\nonumber \\
 & = \frac{d}{d\epsilon}\int yp_{Y,\epsilon}(y\mid m,z,a,w,s=1)\hat{g}_{M\mid a^{*},W,s}\left(m\mid w\right)p_{Z}\left(z\mid a,w,s=0\right)p_{W \mid S}\left(w\mid s=0\right)dv(y,m,z,w)\biggr\vert_{\epsilon=0} \label{derivT}\\
 &  +\frac{d}{d\epsilon}\int yp_{Y}(y\mid m,z,a,w,s=1)\hat{g}_{M\mid a^{*},W,s}\left(m\mid w\right)p_{Z,\epsilon}\left(z\mid a,w,s=0\right)p_{W \mid S}\left(w\mid s=0\right)dv(y,m,a,z,w)\biggr\vert_{\epsilon=0}\nonumber \\
 &  +\frac{d}{d\epsilon}\int yp_{Y}(y\mid m,z,a,w,s=1)\hat{g}_{M\mid a^{*},W,s}\left(m\mid w\right)p_{Z}\left(z\mid a,w,s=0\right)p_{W\mid S\epsilon}\left(w\mid s=0\right)dv(y,m,z,w)\biggr\vert_{\epsilon=0}\nonumber 
\end{align}
\end{scriptsize}\

The first term in \ref{derivT}:

\begin{scriptsize}
\begin{align*}
 & \frac{d}{d\epsilon}\int yp_{Y,\epsilon}(y\mid m,z,a,w,s=1)\hat{g}_{M\mid a^{*},W,s}\left(m\mid w\right)p_{Z}\left(z\mid x=a,w,s=0\right)p_{W \mid S}\left(w\mid s=0\right)dv(y,m,z,w)\biggr\vert_{\epsilon=0}\protect\\
= & \int y\frac{d}{d\epsilon}p_{Y,\epsilon}((y\times s)\mid m,z,x,w,s)\biggr\vert_{\epsilon=0}\hat{g}_{M\mid a^{*},W,s}\left(m\mid w\right)\frac{g_{M}\left(m\mid z,x,w,s\right)}{g_{M}\left(m\mid z,x,w,s\right)}p_{Z}\left(z\mid x=a,w,s=0\right)\frac{p_{Z}\left(z\mid x,w,s\right)}{p_{Z}\left(z\mid x,w,s\right)}\protect\\
 & *\frac{I(s=1,x=a)}{p_{A}\left(x\mid w,s\right)}p_{W \mid S}\left(w\mid s=0\right)\frac{p_{W \mid S}\left(w\mid s\right)}{p_{W \mid S}\left(w\mid s\right)}\frac{p_{S}(s)}{p_{S}(s=1)}dv(y,m,z,x,w,s)\protect\\
\protect\overset{(\ref{eq:3.51})}{=} & \int y\left(\gamma(o)-\mathbb{E}\left[\gamma(o)\mid m,z,x,w,s\right]\right)p_{Y}(ys\mid m,z,x,w,s)\hat{g}_{M\mid a^{*},W,S}\left(m\mid w\right)\frac{g_{M}\left(m\mid z,x,w,s\right)}{g_{M}\left(m\mid z,x,w,s\right)}p_{Z}\left(z\mid x=a,w,s=0\right)\protect\\
 & *\frac{p_{Z}\left(z\mid x,w,s\right)}{p_{Z}\left(z\mid x,w,s\right)}\frac{I(s=1,x=a)p_{A}\left(x\mid w,s\right)}{p_{A}\left(x\mid w,s\right)P_{S}(s=1)}p_{W \mid S}\left(w\mid s=0\right)\frac{p_{W \mid S}\left(w\mid s\right)}{p_{W \mid S}\left(w\mid s\right)}p_{S}(s)dv(y,m,z,x,w,s)\protect\\
= & \int \gamma(o)\biggr(y-\mathbb{E}\biggr[y\mid m,z,x,w,s\biggr]\biggr)\times\\
& \frac{\hat{g}_{M\mid a^{*},W,s}(m\mid w)p_{Z}(z\mid a,w,s=0)p_{S \mid W}(s=0\mid w)I(s=1,x=a)}{g_{M}(m\mid z,x,w,s=1)p_{Z}(z\mid x=a,w,s=1)p_{A}(a\mid w,s=1)p_{S \mid W}(s=1\mid w)p_{S}(s=0)}dP(o)\protect\\
= & \left\langle \gamma,D_{Y}^{*}(P)\right\rangle _{L_{0}^{2}(P)}
\end{align*}
\end{scriptsize}

where 
\begin{scriptsize}
\[
D_{Y}^{*}(P)(O)=\left(Y-\mathbb{E}\left[Y\mid M,Z,A,W\right]\right)\frac{\hat{g}_{M\mid a^{*},W,s}\left(M\mid W\right)p_{Z}\left(Z\mid A,W,S=0\right)p_{S \mid W}\left(S=0\mid W\right)I(S=1,A=a)}{g_{M}\left(M\mid Z,A,W,S\right)p_{Z}\left(Z\mid A,W,S\right)p_{A}\left(A\mid W,S\right)p_{S \mid W}\left(S\mid W\right)P_{S}(S=0)}
\]
\end{scriptsize}

The reader may notice $D^*_Y(P)(O)$ is not a mean 0 function of $Y \mid M, Z, W$ because it also depends on the variable, $A$. Hence, it is not an element of the tangent space under the restricted model, $\mathcal{M}$, that incorporates the instrumental variable exclusion restrictions as described in Section 2 of the main text, which is necessarily a proper subspace of  $L^2(P)$ for $P \in \mathcal{M}^{np}$. Therefore, $D^*(P)(O)$ has an extra orthogonal component in addition to the efficient influence curve under $\mathcal{M}$, making its variance necessarily bigger than the lower bound under model $\mathcal{M}$. This is why the estimators using $D^*(P)(O)$ (the EE and TMLE estimators) are not asymptotically efficient for model, $\mathcal{M}$ but are so for $\mathcal{M}^{np}$.\\

The second term in \ref{derivT}:

\begin{scriptsize}
\begin{align*}
 & \frac{d}{d\epsilon}\int yp_{Y}(y\mid m,z,x,w,s=1)\hat{g}_{M\mid a^{*},W,s}\left(m\mid w\right)p_{Z,\epsilon}\left(z\mid a,w,s=0\right)p_{W \mid S}\left(w\mid s=0\right)dv(y,m,z,w)\biggr\vert_{\epsilon=0}\protect\\
= & \int yp_{Y}(y\mid m,z,x,w)\hat{g}_{M\mid a^{*},W,s}\left(m\mid w\right)\frac{d}{d\epsilon}p_{Z,\epsilon}\left(z\mid x,w,s\right)\biggr\vert_{\epsilon=0}\frac{I(s=0)I(x=a)}{p_{A}\left(x\mid w,s\right)p_{S}(s=0)}\protect\\
 & *p_{A}\left(x\mid w,s\right)p_{W \mid S}\left(w\mid s\right)p_{S}(s)dv(y,m,z,x,w,s)\protect\\
\protect\overset{\ref{eq:3.52}}{=} & \int yp_{Y}(y\mid m,x,z,w)\hat{g}_{M\mid a^{*},W,s}\left(m\mid w\right)\left(\mathbb{E}\left[\gamma(o)\mid z,x,w,s\right]-\mathbb{E}\left[\gamma(o)\mid x,w,s\right]\right)p_{Z}(z\mid x,w,s)\protect\\
 & *\frac{I(s=0)I(x=a)}{p_{A}\left(x\mid w,s\right)p_{S}(s=0)}p_{A}\left(x\mid w,s\right)p_{W \mid S}\left(w\mid s\right)p_{S}(s)dv(y,m,z,x,w,s)\protect\\
= & \int \gamma(o)\biggr(\mathbb{E}_{\hat{g}_{M\mid a^{*},W,s}}\biggr(\mathbb{E}\biggr[Y\mid M,A,Z,W\biggr]\mid z,x,w,s=1\biggr)-\\
&\mathbb{E}_{P_{Z\mid A,W,S}}\biggr[\mathbb{E}_{\hat{g}_{M\mid a^{*},W,s}}\biggr(\mathbb{E}\biggr[Y\mid M,A,Z,W\biggr]\mid Z,A,W,S=1\biggr)\mid x,w,s\biggr]\biggr)*\frac{I(s=0,x=a)}{p_{A}\left(x\mid w,s\right)p_{S}(s=0)}dP(o)\protect\\
= & \left\langle \gamma,D_{Z}^{*}(P)\right\rangle _{L_{0}^{2}(P)}
\end{align*}

\end{scriptsize}

We substitute 
\begin{align*}
\bar{Q}_{M}(z,x,w) & =\mathbb{E}_{\hat{g}_{M\mid a^{*},W,s}}\left(\mathbb{E}\left[Y\mid M,A,Z,W\right]\mid z,x,w,s=1\right)\\
\bar{Q}_{Z}(x,w,s) & =\mathbb{E}_{P_{Z\mid A,W,S}}\left[\mathbb{E}_{\hat{g}_{M\mid a^{*},W,s}}\bar{Q}_{M}(Z,A,W)\mid x,w,s\right]
\end{align*}
 and since $x$ represents the treatment, $A$, in the integrals above, we get

$D_{Z}^{*}(P)(O)=\left(\bar{Q}_{M}(Z,A,W)-\bar{Q}_{Z}(A, W, S)\right)\frac{I(S=0,A=a)}{p_{A}\left(A\mid W,S\right)p_{S}(S=0)}.$\\

The third term in \ref{derivT}:
\begin{scriptsize}
\begin{align*}
 & \frac{d}{d\epsilon}\int yp_{Y}(y\mid m,z,a,w,s=1)\hat{g}_{M\mid a^{*},W,s}\left(m\mid w\right)p_{Z}\left(z\mid a,w,s=0\right)p_{W \mid S,\epsilon}\left(w\mid s=0\right)dv(y,m,z,w)\biggr\vert_{\epsilon=0}\protect\\
= & \int yp_{Y}(y\mid m,a,z,w)\hat{g}_{M\mid a^{*},W,s}\left(m\mid w\right)p_{Z}\left(z\mid a,w,s\right)\frac{d}{d\epsilon}p_{W \mid S,\epsilon}\left(w\mid s\right)\biggr\vert_{\epsilon=0}\frac{I(s=0)}{p_{S}(s=0)}p_{S}(s)dv(y,m,z,x,w,s)\protect\\
\protect\overset{\ref{eq:3.53}}{=} & \int yp_{Y}(y\mid m,a,z,w)\hat{g}_{M\mid a^{*},W,s}\left(m\mid w\right)p_{Z}\left(z\mid a,w,s\right)\left(\mathbb{E}\left[\gamma(o)\mid w,s\right]-\mathbb{E}\left[\gamma(o)\mid s\right]\right)\protect\\
 & *p_{W \mid S}(w\mid s)\frac{I(s=0)}{p_{S}(s=0)}p_{S}(s)dv(y,m,z,x,w,s)\protect\\
= & \int S(o)\left(\bar{Q}_{Z}(x=a,w,s)-\Psi(P)\right)\frac{I(s=0)}{p_{S}(s=0)}dP(o)\protect\\
= & \left\langle \gamma,D_{W}^{*}\right\rangle _{L_{0}^{2}(P)}
\end{align*}
\end{scriptsize}

where $D_{W}^{*}(P)(O)=\left(\bar{Q}_{Z}(A=a, W, S)-\Psi(P)\right)\frac{I(S=0)}{p_{S}(S=0)}$

Thus the efficient influence curve is the sum of its orthogonal components:

\[
D^{*}(P)(O)=D_{Y}^{*}(P)(O)+D_{Z}^{*}(P)(O)+D_{W}^{*}(P)(O)
\]

\subsubsection{Derivation of efficient influence curve for model with exclusion restrictions}
\label{eicrestrict}
Next, we derive the EIC for the model that restricts $A$ to affect $M$ and $Y$ only through $Z$. Similar to the above subsection, we allow our treatment mechanism not to be known. 

\begin{thm}
\label{ICSDEr}
The efficient influence curve for our restricted model, where $M$ and $Y$ do not depend directly on $A$, is given by 
\[
D^{*}(P)(O)=D_{Y,r}^{*}(P)(O)+D_{Z}^{*}(P)(O)+D_{W}^{*}(P)(O)
\]
where
\begin{footnotesize}
\[
D_{Y,r}^{*}(P)(O) = \biggr(y-\mathbb{E}\biggr[y\mid m,z,w\biggr]\biggr) \frac{\hat{g}_{M\mid a^{*},W,s}(m\mid w)p_{Z}(z\mid a_0,w,s=0)p_{S \mid W}(s=0\mid w)I(s=1)}{g_{M,r}(m\mid z,w,s)p_{Z\mid W,S}(z\mid w,s)p_{S \mid W}(s\mid w)p_{S}(s=0)}
\]
\end{footnotesize}
\end{thm}

Proof:

We can note that our only task here is to project $D^*_Y(P)$, our component of the influence curve in $T_Y$, onto the subspace of $T_Y$ given by 

$T_{Y,r} = \overline{\{\gamma : \mathbb{E}(\gamma(O) \mid YS, M, Z, W, S) = 0 , \mathbb{E}\gamma(O)^2 < \infty\}}$.  

$p_{YS,r}$ is the conditional density of $ys$ given $m, z, w$ and $p_{M,r}$ is the conditional density of $m$ given $z, w, s$ in the restricted model, i.e. we don't put the instrument, $a$, in those conditional statements as that is the model assumption. We remind the reader that a "bar" signifies the variable and all past variables as in, $\bar{M} = m, z, x, w, s$.

Notice the following:
\begin{align}
p_{A\mid \bar{YS},r}(x \mid ys, m, z, w, s=1) & = \frac{p_{\bar{A},r}(x, ys, m, z, w, s=1)}{p_{O/A}(ys, m, z, w, s=1)} \nonumber \\
& = \frac{p_{Y,r}(y \mid m, z, w)p_{M,r}(m \mid z, w, s=1)p_{\bar{Z}}(z, x, w, s)}{p_{Y,r}(y \mid m, z, w) p_{M,r}(m \mid z, w, s=1)p_{\bar{Z}}( z, w, s=1)}  \nonumber  \\
& = \frac{p_{\bar{Z}}(z, x, w, s=1)}{p_{\bar{Z}/A}( z, w, s=1)} \label{other1}
\end{align}

\begin{align}
p_{A,YS, r}(x, ys \mid m, z, w, s=1) & = \frac{p_{\bar{Y},r}(ys, x, m, z, w, s=1)}{p_{\bar{M},r}(m, z, w, s=1)} \nonumber \\
& = \frac{p_{Y,r}(y \mid m, z, w)p_{\bar{Z}}(z, x, w, s=1)}{p_{\bar{Z}/A}(z, w, s=1)} \label{other2}
\end{align}

Thus from \ref{other1} and \ref{other2} 
\begin{footnotesize}
\begin{align*}
& \prod(D_Y^* \Vert T_{Y,r})\\
 & = \mathbb{E}(D_Y^*(O) \mid YS, M, Z, W, S) - \mathbb{E}(D_Y^*(O) \mid M, Z, W, S)\\
& = \int \biggr(y-\mathbb{E}\biggr[y\mid m,z,w\biggr]\biggr)\times\\
& \frac{\hat{g}_{M\mid a^{*},W,s}(m\mid w)p_{Z}(z\mid a,w,s=0)p_{S \mid W}(s=0\mid w)I(s=1,x=a)}{g_{M,r}(m\mid z,w,s=1)p_{Z}(z\mid a,w,s=1)g_A(a\mid w,1)p_{S \mid W}(1\mid w)p_{S}(0)} p_{A\mid \bar{YS},r}(x \mid ys, m, z, w, s)d\nu(x)\\
& - \int \biggr(y-\mathbb{E}\biggr[y\mid m,z,w\biggr]\biggr)\times\\
& \frac{\hat{g}_{M\mid a^{*},W,s}(m\mid w)p_{Z}(z\mid a,w,s=0)p_{S \mid W}(s=0\mid w)I(s=1,x=a)}{g_{M,r}(m\mid z,w,s=1)p_{Z}(z\mid a,w,s=1)g_A(a\mid w,1)p_{S \mid W}(1\mid w)p_{S}(0)} p_{A\mid \bar{YS},r}(x ,ys\mid m, z, w, s)d\nu(x, ys)\\
& \text{remembering we are integrating with respect to x and all else is fixed in the first integral}\\
& \text{All is fixed but x and ys in the second integral.  Since I(s=1), ys = 1 and s = 1}\\
& = \int \biggr(y-\mathbb{E}\biggr[y\mid m,z,w\biggr]\biggr)\times\\
& \frac{\hat{g}_{M\mid a^{*},W,s}(m\mid w)p_{Z}(z\mid a,w,s=0)p_{S \mid W}(s=0\mid w)I(s=1,x=a)}{g_{M,r}(m\mid z,w,s=1)p_{Z}(z\mid a,w,s=1)g_A(a\mid w,1)p_{S \mid W}(1\mid w)p_{S}(0)} p_{A\mid \bar{YS},r}(x \mid ys, m, z, w, s=1)d\nu(x)\\
& - \int \biggr(y-\mathbb{E}\biggr[y\mid m,z,w\biggr]\biggr)\times\\
& \frac{\hat{g}_{M\mid a^{*},W,s}(m\mid w)p_{Z}(z\mid a,w,s=0)p_{S \mid W}(s=0\mid w)I(s=1,x=a)}{g_{M,r}(m\mid z,w,s=1)p_{Z}(z\mid a,w,s=1)g_A(a\mid w,1)p_{S \mid W}(1\mid w)p_{S}(0)} p_{A\mid \bar{YS},r}(x ,ys\mid m, z, w, s=1)d\nu(x, ys)\\
&\text{use (\ref{other1}) and (\ref{other2}) for the 1st and 2nd integrals respectively:}\\
& = \int \biggr(y-\mathbb{E}\biggr[y\mid m,z,w\biggr]\biggr)\times\\
& \frac{\hat{g}_{M\mid a^{*},W,s}(m\mid w)p_{Z}(z\mid a,w,s=0)p_{S \mid W}(s=0\mid w)I(s=1,x=a)}{g_{M,r}(m\mid z,w,s=1)p_{Z}(z\mid a,w,s=1)g_A(a\mid w,1)p_{S \mid W}(1\mid w)p_{S}(0)} \frac{p_{\bar{Z}}(z, x, w, s=1)}{p_{\bar{Z}}( z, w, s=1)}d\nu(x)\\
& - \underbrace{\int \biggr(y-\mathbb{E}\biggr[y\mid m,z,w\biggr]\biggr)p_{Y,r}(y \mid m, z, w)d\nu(y)}_{\text{is 0}}\times \\
& \int \frac{\hat{g}_{M\mid a^{*},W,s}(m\mid w)p_{Z}(z\mid a,w,s=0)p_{S \mid W}(s=0\mid w)I(s=1,x=a)}{g_{M,r}(m\mid z,w,s=1)p_{Z}(z\mid a,w,s=1)g_A(a\mid w,1)p_{S \mid W}(1\mid w)p_{S}(0)} \frac{p_{\bar{Z}}(z, x, w, s=1)}{p_{\bar{Z}/A}(z, w, s=1)} d\nu(x)\\
& = \biggr(y-\mathbb{E}\biggr[y\mid m,z,w\biggr]\biggr) \frac{\hat{g}_{M\mid a^{*},W,s}(m\mid w)p_{Z}(z\mid a,w,s=0)p_{S \mid W}(s=0\mid w)I(s=1)}{g_{M,r}(m\mid z,w,s)p_{Z \mid W,S}(z\mid w,s)p_{S \mid W}(s\mid w)p_{S}(s=0)}
\end{align*}
\end{footnotesize}
And the proof is complete since the other components of the unrestricted model's influence curve will remain the same.  The reader may note that $p_{Z \mid W,S}(z\mid w,s) = p_{Z}(z\mid 1, w,s) g_A(1 \mid w,s)+ p_{Z}(z\mid 0,w,s)g_A(0 \mid w,s)$, so we need not perform any additional regressions for this restricted model.  

\subsubsection{Robustness}
\label{R2SDE}
Here we derive the remainder term for the EE estimator. We first form an initial estimate 

$$\hat{\Psi}_n^0 = \sum_{i=1}^n \frac{\mathbb{I}(S_i=0)}{\sum_{i=1}^n\mathbb{I}(S=0)} \bar{Q}_{L_0,N}^{0}(A_i=a,W_i,S_i)$$.  

Then we update this estimate by adding the empirical mean of the approximated influence curve.  If we call our approximated influence curve, $\hat{D}_n^{EE}$ our estimating equation (EE) estimate is given by

$$\hat{\Psi}_n^1 = \hat{\Psi}_n^0 + \sum_{i=1}^n\hat{D}_n^{EE}(O_i)$$  This then leads to a second order expansion

\[
\hat{\Psi}_n^1 - \Psi(P_0) = (P_n - P_0)\hat{D}_n^{EE}(O) + R_2(P_n, P_0)
\]

where $R_2(P_n, P_0) = \hat{\Psi}_n^0 - \Psi(P_0) + P_0\hat{D}_n^{EE}(O)$.

The behavior of $R_2$ determines the robustness of our estimator and conditions under which we can guarantee consistency and efficiency \citep{Laan:2006aa, Laan:2011aa}.  

\begin{thm}
\begin{align*}
R_{2}(P_n, P_0) &= \int\left(\bar{Q}_{Y,0}(m,z,w)-\bar{Q}_{Y}(m,z,w)\right)*\\
&\hspace{.2cm} \frac{f_{1,0}f_{2,0}f_{3}f_{4,0}f_{5,0}f_{6}f_7(o) -f_{1}f_{2}f_{3,0}f_{4}f_{5,0}f_{6}f_{7,0}(o)}{g(o)} dv(o)\\
&= \int\left(\bar{Q}_{Y}(m,z,w)-\bar{Q}_{Y,0}(m,z,w)\right)\sum_{i=1}^6 (f_{i,0}-f_{i})(o)h_i(o)\\
& \overset{Cauchy-Schwarz}{\leq}  k\sum_{i=1}^6\Vert \bar{Q}_{Y}-\bar{Q}_{Y,0}\Vert_{L^2(P_0)}\Vert f_{i,0}-f_i\Vert_{L^2(P_0)} 
\end{align*}
Where we substituted the following: $f_{1,0}(o) = g_{M,0}(m \mid m,z,w)$, $f_{2,0}= p_{Z,0}(m \mid a,w,1)$, $f_{3,0}= p_{Z,0}(m \mid a,w,0)$, $f_{4,0}= p_{A,0}(x=a \mid w,1)$, $f_{5,0}= p_{A,0}(x=a \mid w,0)$, $f_{6,0}= p_{S,0}(s=1 \mid w)$ and $f_{7,0}= p_{S,0}(s=0 \mid w)$ and dropping the subscript, 0, indicates the estimated counterpart.  $h_i$ is a bounded function by the positivity assumption (see section \ref{idSDE}) and thus the last inequality holds with a sufficiently large $k$. 
\end{thm}
\begin{scriptsize}
\begin{align*}
R_{2} & = \underbrace{\Psi_{n}-\Psi(P_{0})+P_{0}\left\{ \left(Y-\bar{Q_{Y}}(M,Z,W\right)\frac{\hat{g}_{a^{*},W,s}(M\mid W)p_{Z}(Z\mid A,W,S=0)p_{S \mid W}(S=0\mid W)\mathbb{I}(S=1,A=a)}{g_{M}(M\mid Z,W,S)p_{Z}(Z\mid A,W,S)p_{A}(A\mid W,S)p_{S\mid W}(S\mid W)P(S=0)}\right\}}_{\text{term 1}} \\
 &  + \underbrace{P_{0}\left\{ \left(\bar{Q}_{L_1}(Z,A,W)-\bar{Q}_{L_0}(A,W,S)\right)\frac{\mathbb{I}(S=0,A=a)}{p_{A}(A\mid W,S)P(S=0)}\right\} +P_{0}\left\{ \left(\bar{Q}_{L_0}(a,W,S)-\Psi_{n}\right)\frac{\mathbb{I}(S=0)}{P(S=0)}\right\}}_{\text{term 2}} \\
 & \approx  P_{0}\left\{ \left(Y-\bar{Q}(M,Z,W\right)\frac{\hat{g}_{a^{*},W,S}(M\mid W,S=0)p_{Z}(Z\mid A,W,S=0)p_{S \mid W}(S=0\mid W)\mathbb{I}(S=1,A=a)}{g_{M}(M\mid Z,W,S)p_{Z}(Z\mid A,W,S)p_{A}(A\mid W,S)p_{S\mid W}(S\mid W)P(S=0)}\right\} \\
 &  +P_{0}\left\{ \left(\bar{Q}_{L_1}(Z,A,W)-\bar{Q}_{L_0}(A,W,S)\right)\frac{\mathbb{I}(S=0,A=a)}{p_{A}(A\mid W,S)P(S=0)}\right\} +P_{0}\left\{ \left(\bar{Q}_{L_0}(a,W,S)-\Psi_{0}\right)\frac{\mathbb{I}(S=0)}{P(S=0)}\right\} \\
 & = \underbrace{P_{0}\left\{ \left(\bar{Q}_{Y,0}(M,Z,W)-\bar{Q}_{Y}(M,Z,W\right)\frac{\hat{g}_{a^{*},W,S}(M\mid W,S=0)p_{Z}(Z\mid A,W,S=0)p_{S \mid W}(S=0\mid W)\mathbb{I}(S=1,A=a)}{g_{M}(M\mid Z,W,S)p_{Z}(Z\mid A,W,S)p_{A}(A\mid W,S)p_{S\mid W}(S\mid W)P(S=0)}\right\}}_{\text{term 3}} \\
 &  +\underbrace{P_{0}\left\{ \left(\bar{Q}_{L_1,0}(Z,A,W)-\bar{Q}_{L_0}(A,W,S)\right)\frac{\mathbb{I}(S=0,A=a)}{p_{A}(A\mid W,S)P(S=0)}\right\} +P_{0}\left\{ \left(\bar{Q}_{L_0}(a,W,S)-\bar{Q}_{L_0,0}(a,W,S)\right)\frac{\mathbb{I}(S=0)}{P(S=0)}\right\}}_{\text{term 4}} \\
 &  +\underbrace{P_{0}\left\{ \left(\bar{Q}_{L_1}(Z,A,W)-\bar{Q}_{L_1,0}(Z,W)\right)\frac{\mathbb{I}(S=0,A=a)}{p_{A}(A\mid W,S)P(S=0)}\right\}}_{\text{term 5}}
\end{align*}
\end{scriptsize}

The treatment mechanism (model for A) being well-specified makes term 4 disappear above. Clearly terms 3 and 5 only cancel if
the models for M, Z and S also are well-specified. 

Integrating terms 3 and 5, we get 
\begin{scriptsize}
\begin{align*}
 &  \int\left(\bar{Q}_{Y,0}(m,Z,W)-\bar{Q}_{Y}(m,Z,W)\right)\hat{g}_{a^{*},W}(m\mid Z)\frac{g_{M,0}(m\mid z,w,1)}{g_{M}(m\mid z,w,1))}\frac{p_{Z,0}(Z\mid a,w,1)}{p_{Z}(Z\mid a,w,1)}p_{Z}(Z\mid a,w,0)\times\\
 &\frac{p_{A,0}(a\mid w,1)}{p_{A}(a\mid w,1)}\frac{p_{S\mid W,0}(1\mid w)p_{S}(0\mid w)}{p_{S \mid W}(1\mid w)P(s=0)}dv(o)\\
 &  +\int\left(\bar{Q}_{Y}(m,z,w)-\bar{Q}_{Y,0}(m,z,w)\right)\hat{g}_{a^{*},W}(m\mid w)p_{Z,0}(Z\mid a,w,0)\frac{p_{A,0}(a\mid w,0)}{p_{A}(a\mid w,0)}\frac{p_{S \mid W,0}(0\mid w)}{P(s=0)}dv(o)
\end{align*}
\end{scriptsize}

If we add fractions and put everything in one integral we have:

\begin{align*}
&\int\left(\bar{Q}_{Y}(m,z,w)-\bar{Q}_{Y,0}(m,z,w)\right)\frac{f_{1,0}f_{2,0}f_{3}f_{4,0}f_5f_{6,0}f_{7}(o) -f_{1}f_{2}f_{3,0}f_{4}f_{5,0}f_{6}f_{7,0}(o)}{g(o)} dv(o)\\
&= \int\left(\bar{Q}_{Y}(m,z,w)-\bar{Q}_{Y,0}(m,z,w)\right)\sum_{i=1}^6 (f_{i,0}-f_{i})(o)h_i(o)\\
& \overset{Cauchy-Schwarz}{\leq}  k\sum_{i=1}^6\Vert \bar{Q}_{Y}-\bar{Q}_{Y,0}\Vert_{L^2(P_0)}\Vert f_{i,0}-f_i\Vert_{L^2(P_0)} 
\end{align*}

$h_i$ is a bounded function by the positivity assumption (see section \ref{idSDE}) and thus the last inequality holds with a sufficiently large $k$, completing the proof\\

We conclude that a product of $L^2(P_0)$ norms between the bias in estimating the outcome model and the bias of the regression models for $M \mid Z,W,S$, $Z \mid A,W,S$, $A \mid W,S$ and $S\mid W$ must be such that the product converges to 0 in probability when multiplied by $\sqrt{n}$.  Also, $\sqrt{n}\Vert \bar{Q}_{L_1}-\bar{Q}_{L_1,0} \Vert_{L^2{P_0}} \Vert p_{A}-p_{A,0}\Vert_{L^2{P_0}}$ must also converge to 0 in probability for the TMLE to be consistent and asymptotically linear with limiting variance that of the TMLE influence curve (efficient for the broader model including dependence of $Y$ and $M$ on $A$).  Such conditions are guaranteed asymptotically when using the highly adaptive lasso to fit the regressions if the true regressions are of finite sectional variation norm and are left-hand continuous with right-hand limits \citep{Laan:2015aa}.\\

To summarize, if the outcome model is misspecified then we need to correctly specify all the models $M \mid Z,W,S$,  $Z \mid A,W,S$, $A \mid W,S$ and $S\mid W$ for the EE estimator to be consistent and asymptotically efficient.  If we specify the outcome model correctly so that  $\sqrt{n}\Vert \bar{Q}_{Y}-\bar{Q}_{Y,0}\Vert_{L^2(P_0)}$ converges to 0 in probability, then clearly term 1 disappears by the Cauchy-Schwarz inequality, and we have to consider term 2, which disappears if $\sqrt{n}\Vert \bar{Q}_{L_1}-\bar{Q}_{L_1,0} \Vert_{L^2{P_0}} \Vert p_{A}-p_{A,0}\Vert_{L^2{P_0}}$ converges to 0 in probability.  Thus if we correctly specify the regression, $\bar{Q}_{L_1}(Z,W) \mid A, W, S $, along with the regression for the outcome, then we need not specify any other model correctly to obtain a consistent estimator.  
 If we correctly specify the outcome and misspecify the second regression, $\bar{Q}_{L_1} \mid A, W, S $, then we need to correctly specify the treatment mechanism in order to be consistent.  

\subsection{Targeted minimum loss-based estimator}
\subsubsection{Derivation of efficient influence curve}
The derivations of the EICs for the less restricted and more restricted models are given in subsections \ref{eicunrestrict} and \ref{eicrestrict}. 

\subsubsection{Robustness}
For our TMLE estimator:
$\Psi(P_n^*)-\Psi(P_0) = (P_n-P_0)D^*(P_n^*)+R_2(P_n^*,P_0)$ where $P_0$ is the true observed data generating distribution and $P_n^*$ is the TMLE updated estimation of $P_0$.  Since the empirical mean of $D^*(P_n^*)=0$ by virtue of the TMLE algorithm, we have $R_2(P_n^*,P_0)=\Psi(P_n^*)-\Psi(P_0)+P_0D^*(P_n^*)$.  The robustness properties of the TMLE estimator are identical to the EE estimator, see Subsection \ref{R2SDE}.

\section{Simulation}

\begin{table}[H]
\centering
\footnotesize
\caption{Simulation data-generating mechanisms.}
\begin{tabular}{| p{2.5cm} | p{13.0cm} | }
  \hline
\multicolumn{2}{|c|}{Data Generating Mechanism 1}\\
\hline 
  $W_1 \sim bernoulli $ & $P(W_1 = 1) = 0.5$ \\
$W_2 \sim bernoulli $ & $P(W_2 = 1) = expit(0.4 + 0.2W_1)$\\
$S \sim bernoulli $& $P(S = 1) = expit(3W_2 - 1)$\\
$A \sim bernoulli $& $P(A = 1) = 0.5 $ \\
$Z \sim bernoulli $& $P(Z = 1) = expit(-3A + -0.2S + 2W_2 + 0.2AW_2 - 
        0.2AS + 0.2W_2S + 2AW_2S - 0.2)$\\
$M \sim bernoulli $& $P(M = 1) = expit(1Z + 6W_2Z - 2W_2 - 2)$\\
$Y \sim bernoulli $& $P(Y = 1) = expit(log(1.2) + log(40)Z - log(30)M - log(1.2)
        W_2 - log(40)W_2Z)$\\
        \hline
\multicolumn{2}{|c|}{Data Generating Mechanism 2}\\       
\hline
$W_1 \sim bernoulli $ & $P(W_1 = 1) = 0.5$ \\
$W_2 \sim bernoulli $ & $P(W_2 = 1) = expit(0.4 + 0.2W_1)$\\
$S \sim bernoulli $& $P(S = 1) = expit(3W_2 - 1)$\\
$A \sim bernoulli $& $P(A = 1) = 0.5 $ \\
$Z \sim bernoulli $& $P(Z = 1) = expit(-0.1 A + -0.2 S + 0.2 W_2 + 5 AW_2 + 
        0.14 A S + 0.2 W_2 S - 0.2 A W_2 S - 1)$\\
$M \sim bernoulli $& $P(M = 1) = expit(1Z + 3ZW_2 + 0.2ZS - 0.2
        W_2S + 2W_2Z + 0.2S - 0.2ZW_2S - W_2 - 
        2)$\\
$Y \sim bernoulli $& $P(Y = 1) = expit(-6Z + 0.2ZW_2 + 2ZM + 2W_2
        M - 2W_2 + 4M + 1ZW_2M - 0.2)$\\
   \hline
   \multicolumn{2}{|c|}{Data Generating Mechanism 3}\\       
\hline
$W_1 \sim bernoulli $ & $P(W_1 = 1) = 0.5$ \\
$W_2 \sim bernoulli $ & $P(W_2 = 1) = expit(0.4 + 0.2W_1)$\\
$S \sim bernoulli $& $P(S = 1) = expit(3W_2 - 1)$\\
$A \sim bernoulli $& $P(A = 1) = 0.5 $ \\
$Z \sim bernoulli $& $P(Z = 1) = expit(-3  A + 2  S + 2  W_2 + 0.2  A  W_2 - 
        0.2  A  S + 0.2  W_2  S + 2  A  W_2  S - 0.2)$\\
$M \sim bernoulli $& $P(M = 1) = expit(3  Z - 0.2  Z  W_2 + 0.2  Z  S - 0.2  
        W_2  S + 2  W_2  Z + 0.2  S - 0.2  Z  W_2  S - W_2 - 
        2)$\\
$Y \sim bernoulli $& $P(Y = 1) = expit(-6  Z + 0.2  Z  W_2 + 2  Z  M + 2  W_2  
        M - 0.2  W_2 + 4  M + 1  Z  W_2  M - 0.2)$\\
   \hline
\end{tabular}
\end{table}

DGM 1 is intended to break when Y and M models are misspecified, especially. DGM 2 is intended to break when Y and Z models are misspecified, especially.  DGM 3 is intended to break when Y and S models are misspecified, especially.  \\

\begin{singlespacing}
{\footnotesize
\begin{longtable}{|p{2.2cm}| p{1.1cm} p{1cm} p{1.5cm} p{1cm} p{1.5cm}  p{1cm}  p{2.3cm}  | }

\caption{Simulation results comparing estimators of the transported direct and indirect effects for DGP 1 under well-specified models for 
sample sizes 100, 500 and 5000}
\\
\hline

Estimator & Bias & \multicolumn{2}{ c  }{Efficiency} & \multicolumn{2}{ c  }{95\%CI Cov}  & RMSE & \begin{scriptsize}\% Out of Bds\end{scriptsize} \\  &      & \multicolumn{1}{ l  }{IC}& \multicolumn{1}{ l  }{Bootstrapping}     &  \multicolumn{1}{ l  }{IC}& \multicolumn{1}{ l  }{Bootstrapping}                 &   &  \\   
\hline 
\endfirsthead

\multicolumn{8}{c}%
{{\tablename\ \thetable{} -- continued from previous page}} \\
\hline
Estimator & Bias & \multicolumn{2}{ c  }{Efficiency} & \multicolumn{2}{ c  }{95\%CI Cov}  & RMSE &  \begin{scriptsize}\% Out of Bds\end{scriptsize} \\  &      & \multicolumn{1}{ l  }{IC}& \multicolumn{1}{ l  }{Bootstrapping}     &  \multicolumn{1}{ l  }{IC}& \multicolumn{1}{ l  }{Bootstrapping}                 &   &  \\   
\hline
\endhead

\hline \multicolumn{8}{|r|}{{Continued on next page}} \\ \hline
\endfoot

\hline
\endlastfoot
\multicolumn{8}{| c | }{DGM 1, N=100} \\ \hline
\multicolumn{8}{| c | }{Transport stochastic direct effect } \\ \hline
SDE\_tmle\_eff & 0.015 & 86.77 & 161.49 & 0.807 & 0.959 & 0.114 & 0 \\ 
SDE\_EE\_eff & 0.016 & 86.77 & $129.06$ & 0.812 & 0.919 & 0.112 & 0 \\ 
SDE\_tmle & 0.028 & 201.37 & 292.39 & 0.82 & 0.941 & 0.252 & 0 \\ 
SDE\_EE & 0.024 & 216.18 & 4222.44 & 0.923 & 0.924 & 0.210 & 0 \\ 
SDE\_iptw & 0.039 & 301.76 & 305.12 & 0.906 & 0.906 & 0.273 & 0 \\ 
\hline
\multicolumn{8}{| c | }{Transport stochastic indirect effect} \\
\hline
\end{longtable}
}
\end{singlespacing}

\begin{singlespacing}
{\footnotesize
\begin{longtable}{|p{2.1cm}| p{1.1cm} p{1cm} p{1.5cm} p{1cm} p{1.5cm}  p{1cm}  p{2.4cm}  | }

\caption{Simulation results comparing estimators of the transported direct and indirect effects  for DGP 1 under $Z$, $M$, and $S$ models misspecified. For sample sizes 100, 500 and 5000.}
\\
\hline

Estimator & Bias & \multicolumn{2}{ c  }{Efficiency} & \multicolumn{2}{ c  }{95\%CI Cov}  & RMSE & \begin{scriptsize}\% Out of Bds\end{scriptsize} \\  &      & \multicolumn{1}{ l  }{IC}& \multicolumn{1}{ l  }{Bootstrapping}     &  \multicolumn{1}{ l  }{IC}& \multicolumn{1}{ l  }{Bootstrapping}                 &   &  \\   
\hline 
\endfirsthead

\multicolumn{8}{c}%
{{\tablename\ \thetable{} -- continued from previous page}} \\
\hline
Estimator & Bias & \multicolumn{2}{ c  }{Efficiency} & \multicolumn{2}{ c  }{95\%CI Cov}  & RMSE &  \begin{scriptsize}\% Out of Bds\end{scriptsize} \\  &      & \multicolumn{1}{ l  }{IC}& \multicolumn{1}{ l  }{Bootstrapping}     &  \multicolumn{1}{ l  }{IC}& \multicolumn{1}{ l  }{Bootstrapping}                 &   &  \\   
\hline
\endhead

\hline \multicolumn{8}{|r|}{{Continued on next page}} \\ \hline
\endfoot

\hline
\endlastfoot
\multicolumn{8}{| c | }{DGM 1, N=100} \\ \hline
\multicolumn{8}{| c | }{Transport stochastic direct effect  } \\ \hline
SDE\_tmle\_eff & 0.029 & 60.77 & 125.16 & 0.529 & 0.879 & 0.139 & 0 \\ 
SDE\_EE\_eff & 0.029 & 60.38 & 121.08 & 0.539 & 0.873 & 0.134 & 0 \\ 
SDE\_tmle & 0.026 & 87.39 & 167.61 & 0.609 & 0.922 & 0.177 & 0 \\ 
SDE\_EE & 0.029 & 93.83 & 134.66 & 0.693 & 0.878 & 0.159 & 0 \\ 
SDE\_iptw & 0.027 & 182.41 & 172.73 & 0.867 & 0.836 & 0.209 & 0 \\ 
\hline
\multicolumn{8}{| c | }{Transport stochastic indirect effect } \\
\hline
\end{longtable}
}
\end{singlespacing}

\begin{singlespacing}
{\footnotesize
\begin{longtable}{|p{2.2cm}| p{1.1cm} p{1cm} p{1.5cm} p{1cm} p{1.5cm}  p{1cm}  p{2.3cm}  | }

\caption{Simulation results comparing estimators of the transported direct and indirect effects  for DGP 1 under $Y$ and $S$ models misspecified. For sample sizes 100, 500 and 5000.}
\\
\hline

Estimator & Bias & \multicolumn{2}{ c  }{Efficiency} & \multicolumn{2}{ c  }{95\%CI Cov}  & RMSE & \begin{scriptsize}\% Out of Bds\end{scriptsize} \\  &      & \multicolumn{1}{ l  }{IC}& \multicolumn{1}{ l  }{Bootstrapping}     &  \multicolumn{1}{ l  }{IC}& \multicolumn{1}{ l  }{Bootstrapping}                 &   &  \\   
\hline 
\endfirsthead

\multicolumn{8}{c}%
{{ \tablename\ \thetable{} -- continued from previous page}} \\
\hline
Estimator & Bias & \multicolumn{2}{ c  }{Efficiency} & \multicolumn{2}{ c  }{95\%CI Cov}  & RMSE &  \begin{scriptsize}\% Out of Bds\end{scriptsize} \\  &      & \multicolumn{1}{ l  }{IC}& \multicolumn{1}{ l  }{Bootstrapping}     &  \multicolumn{1}{ l  }{IC}& \multicolumn{1}{ l  }{Bootstrapping}                 &   &  \\   
\hline
\endhead

\hline \multicolumn{8}{|r|}{{Continued on next page}} \\ \hline
\endfoot

\hline
\endlastfoot
\multicolumn{8}{| c | }{ DGM 1, N=100} \\ \hline
\multicolumn{8}{| c | }{Transport stochastic direct effect  } \\ \hline
SDE\_tmle\_eff & 0.196 & 99.74 & 130.55 & 0.348 & 0.496 & 0.229 & 0 \\ 
SDE\_EE\_eff & 0.23 & 119.11 & $103.82$ & 0.275 & 0.230 & 0.254 & 0 \\ 
SDE\_tmle & 0.195 & 143.81 & 234.82 & 0.556 & 0.812 & 0.290 & 0 \\ 
SDE\_EE & 0.226 & 182.88 & $161.11$ & 0.487 & 0.494 & 0.291 & 0.100 \\ 
SDE\_iptw & 0.235 & 195.46 & 242.96 & 0.643 & 0.767 & 0.322 & 0 \\ 
\hline
\multicolumn{8}{| c | }{Transport stochastic indirect effect } \\
\hline
\end{longtable}
}
\end{singlespacing}

\begin{singlespacing}
{\footnotesize
\begin{longtable}{|p{2.2cm}| p{1.1cm} p{1cm} p{1.5cm} p{1cm} p{1.5cm}  p{1cm}  p{2.3cm}  | }

\caption{Simulation results comparing estimators of the transported direct and indirect effects  for DGP 1 under well-specified models under the $Y$ and $Z$ models misspecified. For sample sizes 100, 500 and 5000.}
\\
\hline

Estimator & Bias & \multicolumn{2}{ c  }{Efficiency} & \multicolumn{2}{ c  }{95\%CI Cov}  & RMSE & \begin{scriptsize}\% Out of Bds\end{scriptsize} \\  &      & \multicolumn{1}{ l  }{IC}& \multicolumn{1}{ l  }{Bootstrapping}     &  \multicolumn{1}{ l  }{IC}& \multicolumn{1}{ l  }{Bootstrapping}                 &   &  \\   
\hline 
\endfirsthead

\multicolumn{8}{c}%
{{\tablename\ \thetable{} -- continued from previous page}} \\
\hline
Estimator & Bias & \multicolumn{2}{ c  }{Efficiency} & \multicolumn{2}{ c  }{95\%CI Cov}  & RMSE &  \begin{scriptsize}\% Out of Bds\end{scriptsize} \\  &      & \multicolumn{1}{ l  }{IC}& \multicolumn{1}{ l  }{Bootstrapping}     &  \multicolumn{1}{ l  }{IC}& \multicolumn{1}{ l  }{Bootstrapping}                 &   &  \\   
\hline
\endhead

\hline \multicolumn{8}{|r|}{{Continued on next page}} \\ \hline
\endfoot

\hline
\endlastfoot
\multicolumn{8}{| c | }{ DGM 1, N=100} \\ \hline
\multicolumn{8}{| c | }{Transport stochastic direct effect  } \\ \hline
SDE\_tmle\_eff & 0.195 & 92.71 & 116.32 & 0.395 & 0.524 & 0.222 & 0 \\ 
SDE\_EE\_eff & 0.216 & 113.89 & 117.45 & 0.453 & 0.469 & 0.241 & 0 \\ 
SDE\_tmle & 0.076 & 261.30 & 292.91 & 0.824 & 0.865 & 0.323 & 0 \\ 
SDE\_EE & 0.082 & 326.80 & 293.93 & 0.921 & 0.874 & 0.321 & 0.100 \\ 
SDE\_iptw & 0.040 & 271.53 & 271.20 & 0.892 & 0.889 & 0.281 & 0 \\ 
\hline
\multicolumn{8}{| c | }{Transport stochastic indirect effect } \\
\hline
\end{longtable}
}
\end{singlespacing}

\begin{singlespacing}
{\footnotesize
\begin{longtable}{|p{2.2cm}| p{1.1cm} p{1cm} p{1.5cm} p{1cm} p{1.5cm}  p{1cm}  p{2.3cm}  | }

\caption{Simulation results comparing estimators of the transported direct and indirect effects  for DGP 1 under $Y$ and $M$ models misspecified. For sample sizes 100, 500 and 5000.}
\\
\hline

Estimator & Bias & \multicolumn{2}{ c  }{Efficiency} & \multicolumn{2}{ c  }{95\%CI Cov}  & RMSE & \begin{scriptsize}\% Out of Bds\end{scriptsize} \\  &      & \multicolumn{1}{ l  }{IC}& \multicolumn{1}{ l  }{Bootstrapping}     &  \multicolumn{1}{ l  }{IC}& \multicolumn{1}{ l  }{Bootstrapping}                 &   &  \\   
\hline 
\endfirsthead

\multicolumn{8}{c}%
{{\tablename\ \thetable{} -- continued from previous page}} \\
\hline
Estimator & Bias & \multicolumn{2}{ c  }{Efficiency} & \multicolumn{2}{ c  }{95\%CI Cov}  & RMSE &  \begin{scriptsize}\% Out of Bds\end{scriptsize} \\  &      & \multicolumn{1}{ l  }{IC}& \multicolumn{1}{ l  }{Bootstrapping}     &  \multicolumn{1}{ l  }{IC}& \multicolumn{1}{ l  }{Bootstrapping}                 &   &  \\   
\hline
\endhead

\hline \multicolumn{8}{|r|}{{Continued on next page}} \\ \hline
\endfoot

\hline
\endlastfoot
\multicolumn{8}{| c | }{ DGM 1, N=100} \\ \hline
\multicolumn{8}{| c | }{Transport stochastic direct effect  } \\ \hline
SDE\_tmle\_eff & -0.133 & 245.31 & 239.69 & 0.782 & 0.837 & 0.283 & 0 \\ 
SDE\_EE\_eff & -1.216 & 1,423.80 & 1,532.61 & 0.895 & 0.890 & 1.984 & 51.100 \\ 
SDE\_tmle & -0.213 & 429.80 & 338.75 & 0.718 & 0.790 & 0.418 & 0 \\ 
SDE\_EE & -1.259 & 1,611.01 & $1632.61$ & 0.953 & 0.949 & 2.078 & 56.300 \\ 
SDE\_iptw & -0.187 & 1,098.72 & 320.43 & 0.893 & 0.818 & 0.382 & 0 \\ 
\hline
\multicolumn{8}{| c | }{Transport stochastic indirect effect } \\
\hline
\end{longtable}
}
\end{singlespacing}

\begin{singlespacing}
{\footnotesize
\begin{longtable}{|p{2.2cm}| p{1.1cm} p{1cm} p{1.5cm} p{1cm} p{1.5cm}  p{1cm}  p{2.3cm}  | }

\caption{Simulation results comparing estimators of the transported direct and indirect effects  for DGP 1 under the $Y$ model misspecified. For sample sizes 100, 500 and 5000.}
\\
\hline

Estimator & Bias & \multicolumn{2}{ c  }{Efficiency} & \multicolumn{2}{ c  }{95\%CI Cov}  & RMSE & \begin{scriptsize}\% Out of Bds\end{scriptsize} \\  &      & \multicolumn{1}{ l  }{IC}& \multicolumn{1}{ l  }{Bootstrapping}     &  \multicolumn{1}{ l  }{IC}& \multicolumn{1}{ l  }{Bootstrapping}                 &   &  \\   
\hline 
\endfirsthead

\multicolumn{8}{c}%
{{\tablename\ \thetable{} -- continued from previous page}} \\
\hline
Estimator & Bias & \multicolumn{2}{ c  }{Efficiency} & \multicolumn{2}{ c  }{95\%CI Cov}  & RMSE &  \begin{scriptsize}\% Out of Bds\end{scriptsize} \\  &      & \multicolumn{1}{ l  }{IC}& \multicolumn{1}{ l  }{Bootstrapping}     &  \multicolumn{1}{ l  }{IC}& \multicolumn{1}{ l  }{Bootstrapping}                 &   &  \\   
\hline
\endhead

\hline \multicolumn{8}{|r|}{{Continued on next page}} \\ \hline
\endfoot

\hline
\endlastfoot
\multicolumn{8}{| c | }{ DGM 1, N=100} \\ \hline
\multicolumn{8}{| c | }{Transport stochastic direct effect  } \\ \hline
SDE\_tmle\_eff & 0.045 & 157.49 & 170.41 & 0.872 & 0.893 & 0.143 & 0 \\ 
SDE\_EE\_eff & 0.050 & 232.07 & $170.51$ & 0.894 & 0.872 & 0.149 & 0 \\ 
SDE\_tmle & 0.010 & 282.54 & 314.26 & 0.876 & 0.912 & 0.273 & 0 \\ 
SDE\_EE & 0.022 & 355.18 & $289.81$ & 0.950 & 0.902 & 0.258 & 0 \\ 
SDE\_iptw & 0.014 & 297.66 & 306.32 & 0.897 & 0.909 & 0.267 & 0 \\ 
\hline
\multicolumn{8}{| c | }{Transport stochastic indirect effect } \\
\hline
\end{longtable}
}
\end{singlespacing}

\begin{singlespacing}
{\footnotesize
\begin{longtable}{|p{2.2cm}| p{1.1cm} p{1cm} p{1.5cm} p{1cm} p{1.5cm}  p{1cm}  p{2.3cm}  | }

\caption{Simulation results comparing estimators of the transported direct and indirect effects  for DGP 2 under well-specified models for 
sample sizes 100, 500 and 5000.}
\\
\hline

Estimator & Bias & \multicolumn{2}{ c  }{Efficiency} & \multicolumn{2}{ c  }{95\%CI Cov}  & RMSE & \begin{scriptsize}\% Out of Bds\end{scriptsize} \\  &      & \multicolumn{1}{ l  }{IC}& \multicolumn{1}{ l  }{Bootstrapping}     &  \multicolumn{1}{ l  }{IC}& \multicolumn{1}{ l  }{Bootstrapping}                 &   &  \\   
\hline 
\endfirsthead

\multicolumn{8}{c}%
{{ \tablename\ \thetable{} -- continued from previous page}} \\
\hline
Estimator & Bias & \multicolumn{2}{ c  }{Efficiency} & \multicolumn{2}{ c  }{95\%CI Cov}  & RMSE &  \begin{scriptsize}\% Out of Bds\end{scriptsize} \\  &      & \multicolumn{1}{ l  }{IC}& \multicolumn{1}{ l  }{Bootstrapping}     &  \multicolumn{1}{ l  }{IC}& \multicolumn{1}{ l  }{Bootstrapping}                 &   &  \\   
\hline
\endhead

\hline \multicolumn{8}{|r|}{{Continued on next page}} \\ \hline
\endfoot

\hline
\endlastfoot
\multicolumn{8}{| c | }{DGM 2, N=100} \\ \hline
\multicolumn{8}{| c | }{Transport stochastic direct effect  } \\ \hline
SDE\_tmle\_eff & 0.003 & 102.26 & 139.52 & 0.970 & 0.992 & 0.064 & 0 \\ 
SDE\_EE\_eff & 0.004 & 102.13 & $118.65$ & 0.974 & 0.989 & 0.060 & 0 \\ 
SDE\_tmle & 0.007 & 293.07 & 375.10 & 0.852 & 0.945 & 0.218 & 0 \\ 
SDE\_EE & 0.005 & 316.03 & 292.87 & 0.957 & 0.930 & 0.186 & 0 \\ 
SDE\_iptw & 0.009 & 442.80 & 456.03 & 0.909 & 0.922 & 0.267 & 0 \\ 
\hline
\multicolumn{8}{| c | }{Transport stochastic indirect effect } \\
\hline
\end{longtable}
}
\end{singlespacing}

\begin{singlespacing}
{\footnotesize
\begin{longtable}{|p{2.2cm}| p{1.1cm} p{1cm} p{1.5cm} p{1cm} p{1.5cm}  p{1cm}  p{2.3cm}  | }

\caption{Simulation results comparing estimators of the transported direct and indirect effects  for DGP 2 under $Z$, $M$, and $S$ models misspecified. For sample sizes 100, 500 and 5000.}
\\
\hline

Estimator & Bias & \multicolumn{2}{ c  }{Efficiency} & \multicolumn{2}{ c  }{95\%CI Cov}  & RMSE & \begin{scriptsize}\% Out of Bds\end{scriptsize} \\  &      & \multicolumn{1}{ l  }{IC}& \multicolumn{1}{ l  }{Bootstrapping}     &  \multicolumn{1}{ l  }{IC}& \multicolumn{1}{ l  }{Bootstrapping}                 &   &  \\   
\hline 
\endfirsthead

\multicolumn{8}{c}%
{{ \tablename\ \thetable{} -- continued from previous page}} \\
\hline
Estimator & Bias & \multicolumn{2}{ c  }{Efficiency} & \multicolumn{2}{ c  }{95\%CI Cov}  & RMSE &  \begin{scriptsize}\% Out of Bds\end{scriptsize} \\  &      & \multicolumn{1}{ l  }{IC}& \multicolumn{1}{ l  }{Bootstrapping}     &  \multicolumn{1}{ l  }{IC}& \multicolumn{1}{ l  }{Bootstrapping}                 &   &  \\   
\hline
\endhead

\hline \multicolumn{8}{|r|}{{Continued on next page}} \\ \hline
\endfoot

\hline
\endlastfoot
\multicolumn{8}{| c | }{DGM 2, N=100} \\ \hline
\multicolumn{8}{| c | }{Transport stochastic direct effect  } \\ \hline
SDE\_tmle\_eff & 0.001 & 90.70 & 108.20 & 0.918 & 0.988 & 0.062 & 0 \\ 
SDE\_EE\_eff & 0.001 & 90.72 & 108.25 & 0.918 & 0.988 & 0.062 & 0 \\ 
SDE\_tmle & 0.007 & 191.08 & 235.80 & 0.861 & 0.956 & 0.141 & 0 \\ 
SDE\_EE & 0.004 & 207.43 & 191.55 & 0.957 & 0.951 & 0.121 & 0 \\ 
SDE\_iptw & 0.206 & 367.51 & 375.66 & 0.729 & 0.756 & 0.345 & 0 \\ 
\hline
\multicolumn{8}{| c | }{Transport stochastic indirect effect } \\
\hline
\end{longtable}
}
\end{singlespacing}
\begin{singlespacing}
{\footnotesize
\begin{longtable}{|p{2.2cm}| p{1.1cm} p{1cm} p{1.5cm} p{1cm} p{1.5cm}  p{1cm}  p{2.3cm}  | }

\caption{Simulation results comparing estimators of the transported direct and indirect effects  for DGP 2 under $Y$ and $S$ models misspecified. For sample sizes 100, 500 and 5000.}
\\
\hline

Estimator & Bias & \multicolumn{2}{ c  }{Efficiency} & \multicolumn{2}{ c  }{95\%CI Cov}  & RMSE & \begin{scriptsize}\% Out of Bds\end{scriptsize} \\  &      & \multicolumn{1}{ l  }{IC}& \multicolumn{1}{ l  }{Bootstrapping}     &  \multicolumn{1}{ l  }{IC}& \multicolumn{1}{ l  }{Bootstrapping}                 &   &  \\   
\hline 
\endfirsthead

\multicolumn{8}{c}%
{{\tablename\ \thetable{} -- continued from previous page}} \\
\hline
Estimator & Bias & \multicolumn{2}{ c  }{Efficiency} & \multicolumn{2}{ c  }{95\%CI Cov}  & RMSE &  \begin{scriptsize}\% Out of Bds\end{scriptsize} \\  &      & \multicolumn{1}{ l  }{IC}& \multicolumn{1}{ l  }{Bootstrapping}     &  \multicolumn{1}{ l  }{IC}& \multicolumn{1}{ l  }{Bootstrapping}                 &   &  \\   
\hline
\endhead

\hline \multicolumn{8}{|r|}{{Continued on next page}} \\ \hline
\endfoot

\hline
\endlastfoot
\multicolumn{8}{| c | }{DGM 2, N=100} \\ \hline
\multicolumn{8}{| c | }{Transport stochastic direct effect  } \\ \hline
SDE\_tmle\_eff & -0.048 & 198.25 & 193.84 & 0.903 & 0.921 & 0.146 & 0 \\ 
SDE\_EE\_eff & -0.102 & 315.28 & $200.46$ & 0.946 & 0.875 & 0.181 & 0 \\ 
SDE\_tmle & -0.057 & 317.69 & 365.58 & 0.819 & 0.880 & 0.245 & 0 \\ 
SDE\_EE & -0.102 & 554.59 & $425.87$ & 0.960 & 0.811 & 0.340 & 1.100 \\ 
SDE\_iptw & 0.054 & 332.61 & 378.44 & 0.844 & 0.870 & 0.249 & 0 \\ 
\hline
\multicolumn{8}{| c | }{Transport stochastic indirect effect } \\
\hline
\end{longtable}
}
\end{singlespacing}

\begin{singlespacing}
{\footnotesize
\begin{longtable}{|p{2.2cm}| p{1.1cm} p{1cm} p{1.5cm} p{1cm} p{1.5cm}  p{1cm}  p{2.3cm}  | }

\caption{Simulation results comparing estimators of the transported direct and indirect effects  for DGP 2 under $Y$ and $Z$ models misspecified. For sample sizes 100, 500 and 5000.}
\\
\hline

Estimator & Bias & \multicolumn{2}{ c  }{Efficiency} & \multicolumn{2}{ c  }{95\%CI Cov}  & RMSE & \begin{scriptsize}\% Out of Bds\end{scriptsize} \\  &      & \multicolumn{1}{ l  }{IC}& \multicolumn{1}{ l  }{Bootstrapping}     &  \multicolumn{1}{ l  }{IC}& \multicolumn{1}{ l  }{Bootstrapping}                 &   &  \\   
\hline 
\endfirsthead

\multicolumn{8}{c}%
{{ \tablename\ \thetable{} -- continued from previous page}} \\
\hline
Estimator & Bias & \multicolumn{2}{ c  }{Efficiency} & \multicolumn{2}{ c  }{95\%CI Cov}  & RMSE &  \begin{scriptsize}\% Out of Bds\end{scriptsize} \\  &      & \multicolumn{1}{ l  }{IC}& \multicolumn{1}{ l  }{Bootstrapping}     &  \multicolumn{1}{ l  }{IC}& \multicolumn{1}{ l  }{Bootstrapping}                 &   &  \\   
\hline
\endhead

\hline \multicolumn{8}{|r|}{{Continued on next page}} \\ \hline
\endfoot

\hline
\endlastfoot
\multicolumn{8}{| c | }{DGM 2, N=100} \\ \hline
\multicolumn{8}{| c | }{Transport stochastic direct effect  } \\ \hline
SDE\_tmle\_eff & 0.028 & 73.84 & 84.89 & 0.881 & 0.976 & 0.053 & 0 \\ 
SDE\_EE\_eff & 0.031 & 81.31 & 95.66 & 0.884 & 0.969 & 0.058 & 0 \\ 
SDE\_tmle & 0.146 & 892.01 & 513.89 & 0.914 & 0.847 & 0.360 & 0 \\ 
SDE\_EE & 0.576 & 1,247.94 & 1,341.48 & 0.935 & 0.923 & 0.966 & 24.300 \\ 
SDE\_iptw & 0.105 & 878.55 & 480.40 & 0.945 & 0.856 & 0.326 & 0 \\ 
\hline
\multicolumn{8}{| c | }{Transport stochastic indirect effect } \\
\hline
\end{longtable}
}
\end{singlespacing}

\begin{singlespacing}
{\footnotesize
\begin{longtable}{|p{2.2cm}| p{1.1cm} p{1cm} p{1.5cm} p{1cm} p{1.5cm}  p{1cm}  p{2.3cm}  | }

\caption{Simulation results comparing estimators of the transported direct and indirect effects  for DGP 2 under $Y$ and $M$ models misspecified. For sample sizes 100, 500 and 5000.}
\\
\hline

Estimator & Bias & \multicolumn{2}{ c  }{Efficiency} & \multicolumn{2}{ c  }{95\%CI Cov}  & RMSE & \begin{scriptsize}\% Out of Bds\end{scriptsize} \\  &      & \multicolumn{1}{ l  }{IC}& \multicolumn{1}{ l  }{Bootstrapping}     &  \multicolumn{1}{ l  }{IC}& \multicolumn{1}{ l  }{Bootstrapping}                 &   &  \\   
\hline 
\endfirsthead

\multicolumn{8}{c}%
{{ \tablename\ \thetable{} -- continued from previous page}} \\
\hline
Estimator & Bias & \multicolumn{2}{ c  }{Efficiency} & \multicolumn{2}{ c  }{95\%CI Cov}  & RMSE &  \begin{scriptsize}\% Out of Bds\end{scriptsize} \\  &      & \multicolumn{1}{ l  }{IC}& \multicolumn{1}{ l  }{Bootstrapping}     &  \multicolumn{1}{ l  }{IC}& \multicolumn{1}{ l  }{Bootstrapping}                 &   &  \\   
\hline
\endhead

\hline \multicolumn{8}{|r|}{{Continued on next page}} \\ \hline
\endfoot

\hline
\endlastfoot
\multicolumn{8}{| c | }{DGM 2, N=100} \\ \hline
\multicolumn{8}{| c | }{Transport stochastic direct effect  } \\ \hline
SDE\_tmle\_eff & 0.028 & 122.23 & 210.47 & 0.825 & 0.985 & 0.114 & 0 \\ 
SDE\_EE\_eff & 0.055 & 561.86 & $1131.74$ & 0.905 & 0.998 & 0.724 & 8 \\ 
SDE\_tmle & 0.034 & 797.09 & 552.54 & 0.834 & 0.825 & 0.412 & 0 \\ 
SDE\_EE & 0.026 & 1,816.94 & $1828.34$ & 0.974 & 0.972 & 1.374 & 29.500 \\ 
SDE\_iptw & 0.028 & 639.31 & 509.92 & 0.872 & 0.851 & 0.364 & 0 \\ 
\hline
\multicolumn{8}{| c | }{Transport stochastic indirect effect } \\
\hline
\end{longtable}
}
\end{singlespacing}

\begin{singlespacing}
{\footnotesize
\begin{longtable}{|p{2.2cm}| p{1.1cm} p{1cm} p{1.5cm} p{1cm} p{1.5cm}  p{1cm}  p{2.3cm}  | }

\caption{Simulation results comparing estimators of the transported direct and indirect effects  for DGP 2 under $Y$ model misspecified. For sample sizes 100, 500 and 5000.}
\\
\hline

Estimator & Bias & \multicolumn{2}{ c  }{Efficiency} & \multicolumn{2}{ c  }{95\%CI Cov}  & RMSE & \begin{scriptsize}\% Out of Bds\end{scriptsize} \\  &      & \multicolumn{1}{ l  }{IC}& \multicolumn{1}{ l  }{Bootstrapping}     &  \multicolumn{1}{ l  }{IC}& \multicolumn{1}{ l  }{Bootstrapping}                 &   &  \\   
\hline 
\endfirsthead

\multicolumn{8}{c}%
{{ \tablename\ \thetable{} -- continued from previous page}} \\
\hline
Estimator & Bias & \multicolumn{2}{ c  }{Efficiency} & \multicolumn{2}{ c  }{95\%CI Cov}  & RMSE &  \begin{scriptsize}\% Out of Bds\end{scriptsize} \\  &      & \multicolumn{1}{ l  }{IC}& \multicolumn{1}{ l  }{Bootstrapping}     &  \multicolumn{1}{ l  }{IC}& \multicolumn{1}{ l  }{Bootstrapping}                 &   &  \\   
\hline
\endhead

\hline \multicolumn{8}{|r|}{{Continued on next page}} \\ \hline
\endfoot

\hline
\endlastfoot
\multicolumn{8}{| c | }{DGM 2, N=100} \\ \hline
\multicolumn{8}{| c | }{Transport stochastic direct effect  } \\ \hline
SDE\_tmle\_eff & 0.016 & 117.52 & 135.93 & 0.928 & 0.991 & 0.071 & 0 \\ 
SDE\_EE\_eff & 0.017 & 134.03 & $127.09$ & 0.971 & 0.994 & 0.069 & 0 \\ 
SDE\_tmle & -0.005 & 437.29 & 464.11 & 0.856 & 0.888 & 0.296 & 0 \\ 
SDE\_EE & 0.003 & 510.08 & $415.06$ & 0.971 & 0.887 & 0.270 & 0 \\ 
SDE\_iptw & -0.004 & 437.59 & 450.42 & 0.892 & 0.911 & 0.279 & 0 \\ 
\hline
\multicolumn{8}{| c | }{Transport stochastic indirect effect } \\
\hline
\end{longtable}
}
\end{singlespacing}

\begin{singlespacing}
{\footnotesize
\begin{longtable}{|p{2.2cm}| p{1.1cm} p{1cm} p{1.5cm} p{1cm} p{1.5cm}  p{1cm}  p{2.3cm}  | }

\caption{Simulation results comparing estimators of the transported direct and indirect effects  for DGP 3 under well-specified models for 
sample sizes 100, 500 and 5000.}
\\
\hline

Estimator & Bias & \multicolumn{2}{ c  }{Efficiency} & \multicolumn{2}{ c  }{95\%CI Cov}  & RMSE & \begin{scriptsize}\% Out of Bds\end{scriptsize} \\  &      & \multicolumn{1}{ l  }{IC}& \multicolumn{1}{ l  }{Bootstrapping}     &  \multicolumn{1}{ l  }{IC}& \multicolumn{1}{ l  }{Bootstrapping}                 &   &  \\   
\hline 
\endfirsthead

\multicolumn{8}{c}%
{{ \tablename\ \thetable{} -- continued from previous page}} \\
\hline
Estimator & Bias & \multicolumn{2}{ c  }{Efficiency} & \multicolumn{2}{ c  }{95\%CI Cov}  & RMSE &  \begin{scriptsize}\% Out of Bds\end{scriptsize} \\  &      & \multicolumn{1}{ l  }{IC}& \multicolumn{1}{ l  }{Bootstrapping}     &  \multicolumn{1}{ l  }{IC}& \multicolumn{1}{ l  }{Bootstrapping}                 &   &  \\   
\hline
\endhead

\hline \multicolumn{8}{|r|}{{Continued on next page}} \\ \hline
\endfoot

\hline
\endlastfoot
\multicolumn{8}{| c | }{DGM 3, N=100} \\ \hline
\multicolumn{8}{| c | }{Transport stochastic direct effect  } \\ \hline
SDE\_tmle\_eff & -0.020 & 93.10 & 158.60 & 0.823 & 0.955 & 0.115 & 0 \\ 
SDE\_EE\_eff & -0.017 & 93.22 & $113.95$ & 0.828 & 0.896 & 0.099 & 0 \\ 
SDE\_tmle & -0.030 & 126.80 & 239.56 & 0.809 & 0.951 & 0.199 & 0 \\ 
SDE\_EE & -0.020 & 139.54 & $141.15$ & 0.890 & 0.875 & 0.140 & 0 \\ 
SDE\_iptw & -0.043 & 346.24 & 324.14 & 0.840 & 0.871 & 0.327 & 0 \\ 
\hline
\multicolumn{8}{| c | }{Transport stochastic indirect effect } \\
\hline
\end{longtable}
}
\end{singlespacing}

\begin{singlespacing}
{\footnotesize
\begin{longtable}{|p{2.2cm}| p{1.1cm} p{1cm} p{1.5cm} p{1cm} p{1.5cm}  p{1cm}  p{2.3cm}  | }

\caption{Simulation results comparing estimators of the transported direct and indirect effects  for DGP 3 under $Z$, $M$, and $S$ models misspecified. For sample sizes 100, 500 and 5000.}
\\
\hline

Estimator & Bias & \multicolumn{2}{ c  }{Efficiency} & \multicolumn{2}{ c  }{95\%CI Cov}  & RMSE & \begin{scriptsize}\% Out of Bds\end{scriptsize} \\  
&      & \multicolumn{1}{ l  }{IC}& \multicolumn{1}{ l  }{Bootstrapping}     &  \multicolumn{1}{ l  }{IC}& \multicolumn{1}{ l  }{Bootstrapping}                 &   &  \\   
\hline 
\endfirsthead

\multicolumn{8}{c}%
{{ \tablename\ \thetable{} -- continued from previous page}} \\
\hline
Estimator & Bias & \multicolumn{2}{ c  }{Efficiency} & \multicolumn{2}{ c  }{95\%CI Cov}  & RMSE &  \begin{scriptsize}\% Out of Bds\end{scriptsize} \\  &      & \multicolumn{1}{ l  }{IC}& \multicolumn{1}{ l  }{Bootstrapping}     &  \multicolumn{1}{ l  }{IC}& \multicolumn{1}{ l  }{Bootstrapping}                 &   &  \\   
\hline
\endhead

\hline \multicolumn{8}{|r|}{{Continued on next page}} \\ \hline
\endfoot

\hline

\endlastfoot
\multicolumn{8}{| c | }{DGM 3, N=100} \\ \hline
\multicolumn{8}{| c | }{Transport stochastic direct effect  } \\ \hline
SDE\_tmle\_eff & -0.013 & 71.12 & 119.65 & 0.744 & 0.896 & 0.097 & 0 \\ 
SDE\_EE\_eff & -0.013 & 71.11 & 119.59 & 0.745 & 0.896 & 0.097 & 0 \\ 
SDE\_tmle & -0.011 & 95.20 & 159.52 & 0.781 & 0.942 & 0.124 & 0 \\ 
SDE\_EE & -0.012 & 95.46 & 131.92 & 0.844 & 0.920 & 0.106 & 0 \\ 
SDE\_iptw & -0.110 & 196.15 & 206.68 & 0.877 & 0.889 & 0.200 & 0 \\ 
\hline
\multicolumn{8}{| c | }{Transport stochastic indirect effect } \\
\hline
\end{longtable}
}
\end{singlespacing}

\begin{singlespacing}
{\footnotesize
\begin{longtable}{|p{2.2cm}| p{1.1cm} p{1cm} p{1.5cm} p{1cm} p{1.5cm}  p{1cm}  p{2.3cm}  | }

\caption{Simulation results comparing estimators of the transported direct and indirect effects  for DGP 3 under $Y$ and $S$ models misspecified. For sample sizes 100, 500 and 5000.}
\\
\hline

Estimator & Bias & \multicolumn{2}{ c  }{Efficiency} & \multicolumn{2}{ c  }{95\%CI Cov}  & RMSE & \begin{scriptsize}\% Out of Bds\end{scriptsize} \\  
&      & \multicolumn{1}{ l  }{IC}& \multicolumn{1}{ l  }{Bootstrapping}     &  \multicolumn{1}{ l  }{IC}& \multicolumn{1}{ l  }{Bootstrapping}                 &   &  \\   
\hline 
\endfirsthead

\multicolumn{8}{c}%
{{ \tablename\ \thetable{} -- continued from previous page}} \\
\hline
Estimator & Bias & \multicolumn{2}{ c  }{Efficiency} & \multicolumn{2}{ c  }{95\%CI Cov}  & RMSE &  \begin{scriptsize}\% Out of Bds\end{scriptsize} \\  &      & \multicolumn{1}{ l  }{IC}& \multicolumn{1}{ l  }{Bootstrapping}     &  \multicolumn{1}{ l  }{IC}& \multicolumn{1}{ l  }{Bootstrapping}                 &   &  \\   
\hline
\endhead

\hline \multicolumn{8}{|r|}{{Continued on next page}} \\ \hline
\endfoot

\hline
\endlastfoot
\multicolumn{8}{| c | }{DGM 3, N=100} \\ \hline
\multicolumn{8}{| c | }{Transport stochastic direct effect  } \\ \hline
SDE\_tmle\_eff & -0.213 & 74.55 & 170.93 & 0.230 & 0.654 & 0.266 & 0 \\ 
SDE\_EE\_eff & -0.254 & 88.47 & $125.74$ & 0.115 & 0.274 & 0.279 & 0 \\ 
SDE\_tmle & -0.180 & 119.92 & 267.80 & 0.519 & 0.819 & 0.297 & 0 \\ 
SDE\_EE & -0.238 & 150.22 & $177.67$ & 0.453 & 0.574 & 0.284 & 0 \\ 
SDE\_iptw & -0.210 & 308.35 & 290.33 & 0.788 & 0.862 & 0.328 & 0 \\ 
\hline
\multicolumn{8}{| c | }{Transport stochastic indirect effect } \\
\hline
\end{longtable}
}
\end{singlespacing}

\begin{singlespacing}
{\footnotesize
\begin{longtable}{|p{2.2cm}| p{1.1cm} p{1cm} p{1.5cm} p{1cm} p{1.5cm}  p{1cm}  p{2.3cm}  | }

\caption{Simulation results comparing estimators of the transported direct and indirect effects  for DGP 3 under $Y$ and $Z$ models misspecified. For sample sizes 100, 500 and 5000.}
\\
\hline

Estimator & Bias & \multicolumn{2}{ c  }{Efficiency} & \multicolumn{2}{ c  }{95\%CI Cov}  & RMSE & \begin{scriptsize}\% Out of Bds\end{scriptsize} \\  &      & \multicolumn{1}{ l  }{IC}& \multicolumn{1}{ l  }{Bootstrapping}     &  \multicolumn{1}{ l  }{IC}& \multicolumn{1}{ l  }{Bootstrapping}                 &   &  \\   
\hline 
\endfirsthead

\multicolumn{8}{c}%
{{ \tablename\ \thetable{} -- continued from previous page}} \\
\hline
Estimator & Bias & \multicolumn{2}{ c  }{Efficiency} & \multicolumn{2}{ c  }{95\%CI Cov}  & RMSE &  \begin{scriptsize}\% Out of Bds\end{scriptsize} \\  &      & \multicolumn{1}{ l  }{IC}& \multicolumn{1}{ l  }{Bootstrapping}     &  \multicolumn{1}{ l  }{IC}& \multicolumn{1}{ l  }{Bootstrapping}                 &   &  \\   
\hline
\endhead

\hline \multicolumn{8}{|r|}{{Continued on next page}} \\ \hline
\endfoot

\hline
\endlastfoot
\multicolumn{8}{| c | }{DGM 3, N=100} \\ \hline
\multicolumn{8}{| c | }{Transport stochastic direct effect  } \\ \hline
SDE\_tmle\_eff & -0.155 & 79.58 & 135.64 & 0.422 & 0.658 & 0.195 & 0 \\ 
SDE\_EE\_eff & -0.150 & 117.23 & 136.56 & 0.580 & 0.714 & 0.188 & 0 \\ 
SDE\_tmle & 0.005 & 264.09 & 290.02 & 0.806 & 0.874 & 0.287 & 0 \\ 
SDE\_EE & 0.028 & 380.69 & 299.92 & 0.939 & 0.880 & 0.292 & 0.300 \\ 
SDE\_iptw & 0.002 & 280.33 & 287.94 & 0.867 & 0.883 & 0.272 & 0 \\ 
\hline
\multicolumn{8}{| c | }{Transport stochastic indirect effect } \\
\hline
\end{longtable}
}
\end{singlespacing}

\begin{singlespacing}
{\footnotesize
\begin{longtable}{|p{2.2cm}| p{1.1cm} p{1cm} p{1.5cm} p{1cm} p{1.5cm}  p{1cm}  p{2.3cm}  | }

\caption{Simulation results comparing estimators of the transported direct and indirect effects  for DGP 3 under $Y$ and $M$ models misspecified. For sample sizes 100, 500 and 5000.}
\\
\hline

Estimator & Bias & \multicolumn{2}{ c  }{Efficiency} & \multicolumn{2}{ c  }{95\%CI Cov}  & RMSE & \begin{scriptsize}\% Out of Bds\end{scriptsize} \\  &      & \multicolumn{1}{ l  }{IC}& \multicolumn{1}{ l  }{Bootstrapping}     &  \multicolumn{1}{ l  }{IC}& \multicolumn{1}{ l  }{Bootstrapping}                 &   &  \\   
\hline 
\endfirsthead

\multicolumn{8}{c}%
{{ \tablename\ \thetable{} -- continued from previous page}} \\
\hline
Estimator & Bias & \multicolumn{2}{ c  }{Efficiency} & \multicolumn{2}{ c  }{95\%CI Cov}  & RMSE &  \begin{scriptsize}\% Out of Bds\end{scriptsize} \\  &      & \multicolumn{1}{ l  }{IC}& \multicolumn{1}{ l  }{Bootstrapping}     &  \multicolumn{1}{ l  }{IC}& \multicolumn{1}{ l  }{Bootstrapping}                 &   &  \\   
\hline
\endhead

\hline \multicolumn{8}{|r|}{{Continued on next page}} \\ \hline
\endfoot

\hline
\endlastfoot
\multicolumn{8}{| c | }{DGM 3, N=100} \\ \hline
\multicolumn{8}{| c | }{Transport stochastic direct effect  } \\ \hline
SDE\_tmle\_eff & -0.038 & 130.75 & 184.19 & 0.744 & 0.913 & 0.157 & 0 \\ 
SDE\_EE\_eff & -0.035 & 198.80 & $213.89$ & 0.805 & 0.873 & 0.184 & 0.200 \\ 
SDE\_tmle & 0.055 & 197.50 & 318.16 & 0.574 & 0.859 & 0.338 & 0 \\ 
SDE\_EE & -0.006 & 409.96 & $339.35$ & 0.876 & 0.876 & 0.414 & 2.200 \\ 
SDE\_iptw & 0.020 & 365.53 & 331.20 & 0.822 & 0.860 & 0.325 & 0 \\ 
\hline
\multicolumn{8}{| c | }{Transport stochastic indirect effect } \\
\hline
\end{longtable}
}
\end{singlespacing}

\begin{singlespacing}
{\footnotesize
\begin{longtable}{|p{2.2cm}| p{1.1cm} p{1cm} p{1.5cm} p{1cm} p{1.5cm}  p{1cm}  p{2.3cm}  | }

\caption{Simulation results comparing estimators of the transported direct and indirect effects  for DGP 3 under $Y$ model misspecified.  For sample sizes 100, 500 and 5000.}
\\
\hline

Estimator & Bias & \multicolumn{2}{ c  }{Efficiency} & \multicolumn{2}{ c  }{95\%CI Cov}  & RMSE & \begin{scriptsize}\% Out of Bds\end{scriptsize} \\  &      & \multicolumn{1}{ l  }{IC}& \multicolumn{1}{ l  }{Bootstrapping}     &  \multicolumn{1}{ l  }{IC}& \multicolumn{1}{ l  }{Bootstrapping}                 &   &  \\   
\hline 
\endfirsthead

\multicolumn{8}{c}%
{{ \tablename\ \thetable{} -- continued from previous page}} \\
\hline
Estimator & Bias & \multicolumn{2}{ c  }{Efficiency} & \multicolumn{2}{ c  }{95\%CI Cov}  & RMSE &  \begin{scriptsize}\% Out of Bds\end{scriptsize} \\  &      & \multicolumn{1}{ l  }{IC}& \multicolumn{1}{ l  }{Bootstrapping}     &  \multicolumn{1}{ l  }{IC}& \multicolumn{1}{ l  }{Bootstrapping}                 &   &  \\   
\hline
\endhead

\hline \multicolumn{8}{|r|}{{Continued on next page}} \\ \hline
\endfoot

\hline
\endlastfoot
\multicolumn{8}{| c | }{DGM 3, N=100} \\ \hline
\multicolumn{8}{| c | }{Transport stochastic direct effect  } \\ \hline
SDE\_tmle\_eff & -0.067 & 114.03 & 177.02 & 0.731 & 0.893 & 0.158 & 0 \\ 
SDE\_EE\_eff & -0.095 & 153.08 & $160.44$ & 0.766 & 0.817 & 0.161 & 0 \\ 
SDE\_tmle & 0.019 & 189.07 & 312.17 & 0.598 & 0.876 & 0.314 & 0 \\ 
SDE\_EE & -0.063 & 335.43 & $258.40$ & 0.857 & 0.854 & 0.335 & 1.600 \\ 
SDE\_iptw & -0.015 & 349.83 & 326.79 & 0.868 & 0.883 & 0.306 & 0 \\ 
\hline
\multicolumn{8}{| c | }{Transport stochastic indirect effect } \\
\hline
\end{longtable}
}
\end{singlespacing}

\end{document}